\documentclass[twocolumn]{aa}
\usepackage{graphicx}
\usepackage{txfonts}
\usepackage{enumitem}
\usepackage{ulem}  % for strikethrough
\usepackage{xcolor}

\begin{document}

   \title{Europa's Lyman-$\alpha$ emissions from HST/STIS observations}
   %\subtitle{}

   \author{L. Roth\inst{1}
          \and
          K.D. Retherford\inst{2,3}
          \and
          J. Saur\inst{4}
          \and
          D.F. Strobel\inst{5}         
         \and
          T. Becker\inst{2,3}
          \and
          S. Bergman \inst{1}
          \and
          A. Blöcker \inst{6}
          \and
          S.R. Carberry Mogan \inst{1}
          \and
           C. Grava \inst{2}
          \and
          M. Ivchenko \inst{1}
          \and
          S. Joshi \inst{1}
          \and
           M.A. McGrath \inst{7}
          \and
          F. Nimmo\inst{8}
          \and
          L. Paganini \inst{9}
          \and
          W. Pryor\inst{10}
           \and
          J.R. Spencer\inst{11}
          }

    \institute{KTH Royal Institute of Technology, Stockholm, Sweden\\
    \email{lorenzr@kth.se}
    \and Southwest Research Institute, San Antonio TX, USA 
    \and University of Texas at San Antonio, San Antonio, TX 78249, USA
    \and Institute of Geophysics and Meteorology, University of Cologne, Cologne, Germany
    \and Department of Earth and Planetary Sciences, The Johns Hopkins University, Baltimore, MD, USA
    \and Space Research Institute, Austrian Academy of Sciences, Schmiedlstrasse 6, A-8042 Graz, Austria
    \and SETI Institute, Mountain View, CA, USA
    \and Department of Earth and Planetary Sciences, University of California Santa Cruz, CA, USA
    \and formerly at American University, Washington, DC, USA
    \and Central Arizona College, Coolidge, AZ, USA
    \and Southwest Research Institute, Boulder CO, USA \\
             }

%   \date{Received September 15, 1996; accepted March 16, 1997}

  \abstract
   {An image of Lyman-$\alpha$ (Ly$\alpha$) emission from Europa obtained with the Hubble Space Telescope's Space Telescope Imaging Spectrograph (HST/STIS) provided the first evidence for localized water vapor (H$_2$O) aurora, potentially originating from outgassing. Subsequent STIS observations revealed the presence of a global atomic hydrogen (H) exosphere at Europa.}
   {We present a comprehensive analysis of STIS Ly$\alpha$ observations of Europa acquired in 1999 and between 2012 and 2020 to search for localized auroral emissions and to constrain the properties of Europa’s H exosphere.}
% Methods
   {The complete dataset of the STIS observations obtained when Europa was sunlit and not transiting Jupiter is analyzed. We construct a model that accounts for all known sources of Ly$\alpha$ emission, including resonantly scattered sunlight from Europa’s H exosphere. To identify localized anomalies such as H$_2$O aurora, we subtract the modeled Ly$\alpha$ emission and analyze the residuals.}
    {Emission from Europa’s H exosphere is detected at all observing epochs but is attenuated by absorption in Earth’s exosphere when Europa’s radial velocity relative to Earth, and thus the Doppler shift, is small. From the velocity dependence of this attenuation, we estimate an H-exosphere temperature of $\sim$1000~K and derive an upper limit of 5100~K. For the best-constrained epoch in 2014/2015, we infer a vertical H column density of $1.4 \times 10^{12}$~cm$^{-2}$ and an H source rate of $1.1\times10^{27}$~s$^{-1}$. 

    No localized emission enhancements are detected in any of the observations, including the image previously interpreted as evidence for H$_2$O aurora near Europa’s south pole. The discrepancy with earlier results arises primarily from differences in the assumed position of Europa’s disk on the detector. The inclusion of an H-exosphere signal in the present analysis also contributes to this difference. When adopting the same disk position as in the previous study and neglecting the H-exosphere signal, the localized emission enhancement is again detected with similar statistical significance. However, with the updated approach to disk positioning and the more complete modeling of emission sources, including the H exosphere, we consider the results presented here as the preferred interpretation.}
   {We find evidence for a persistent hydrogen exosphere at Europa but no evidence for localized water vapor.}

   \maketitle
%
%-------------------------------------------------------------------

\section{Introduction}

Jupiter's moon Europa possesses a tenuous atmosphere, which is constantly replenished from weathering of the icy surface \citep{mcgrath09,johnson09}. Irradiation by ions from Jupiter's magnetospheric plasma and radiolysis of the surface water ice is considered to be the main source for the molecular oxygen (O$_2$) and hydrogen (H$_2$) in the atmosphere \citep[e.g.][]{johnson09}, although recent studies suggest that electron-induced radiolysis might also be an important source for icy moon atmospheres \citep{galli2018,meierloeffler2000}. O$_2$ is the globally dominant species near the surface and forms a near-surface layer with an expected scale height of roughly 20–50 km \citep{shematovich05}. The lighter H$_2$ has a much larger scale height of hundreds of kilometers, and tends to more readily escape \citep{smyth06}. Surface ice sublimation can be a source for H$_2$O molecules in the atmosphere \citep{plainaki2018}, but significant fluxes would be expected only in the subsolar region. Water-group species in the atmosphere like OH, O and H can be produced by dissociation of the primary molecular atmosphere but also directly through radiolysis of the ice. The atmospheric density likely varies with local time, hemisphere, and magnetospheric conditions.

Hubble Space Telescope (HST) observations of far-UV atomic oxygen emissions at 1356~Å and 1304~Å have provided the first detection of, and further constraints on, Europa’s molecular atmosphere \citep{hall95,hall98}. The relative intensities of the O I 1304~Å and O I 1356~Å emissions provide a means to distinguish between different oxygen-bearing atmospheric species \citep{Roth2021}. Using HST’s Space Telescope Imaging Spectrograph (STIS), these UV emissions can be spectrally separated and spatially resolved \citep{roth16-eur}.

Near Europa’s limb, the observed oxygen emission ratios (i.e., the ratio of the OI1356~Å to the OI1304~Å emissions) are consistent with an O$_2$-dominated atmosphere. In contrast, in the subsolar region on the trailing hemisphere, the lower OI1356~Å/OI1304~Å ratio suggests that H$_2$O is the dominant species \citep{Roth2021-Eur}. Observations of Europa’s optical aurora are likewise consistent with an O$_2$-dominated atmosphere and provide weak evidence for the presence of H$_2$O as well \citep{dekleer18,dekleer2023}. However, optical observations are limited to disk-integrated quantities during the cooler eclipse phase. In contrast, spatially resolved STIS UV measurements are obtained in sunlight, when warm ice sublimates more readily.

Further indirect constraints on the atmosphere were recently provided by measurements of H$_2^+$, O$_2^+$, and O$^+$ pick-up ions near Europa as well as H$_2^+$ farther away from the moon by the Juno spacecraft \citep{szalay2022,szalay2024}. These measurements provide first evidence for the H$_2$ component and suggest that the turnover (production and loss) of both H$_2$ and O$_2$ in the atmosphere is lower than previous studies assumed \citep[e.g.,][]{smyth06}.

While molecular hydrogen has not been detected yet with remote sensing observations, a widely extended atomic hydrogen exosphere was measured through attenuation at the Ly$\alpha$ line (1216~Å) in HST/STIS observations of Europa in transit of Jupiter \citep{roth17-europa}. The first estimations from \cite{roth17-europa} suggested that this H exosphere is more dilute those that of Ganymede \citep{barth97,feldman00,alday17,naesenius2025} and Callisto \citep{barth1997b,roth17-callisto}, which were detected through Ly$\alpha$ emissions from resonantly scattered sunlight. A comparison of Europa's transit data to later HST observations of Ganymede in transit \citep{Roth2023}, however, indicate that Europa's H exosphere densities were underestimated by \citet{roth17-europa} (in the transit attenuation method): Europa's H abundance should be about five times higher and thus also detectable in emission (not only absorption) in HST Ly$\alpha$ data. In the initial study by \cite{roth17-europa}, the line-center cross section was assumed to be the effective cross section for the attenuation, but a lower, line-integrated cross section should have been used.

The spectral UV images of Europa taken by HST/STIS \citep{mcgrath09,roth16-eur} map emissions at the two oxygen multiplets around 1356~Å and 1304~Å as well as the hydrogen Ly$\alpha$ emission (1216~Å). In one particular observation taken in December 2012 a localized emission surplus was inferred that is consistent with electron-impact dissociative of H$_2$O as the source \citep{roth14-science}. Such a localized abundance of H$_2$O requires a local source because of the short lifetime of water vapor in Europa's atmosphere of some tens of minutes (the time an H$_2$O molecule is aloft before returning to and presumably condensing on the icy surface), and the detection was interpreted as outgassing into a plume. An apparent surplus of OI1304~Å emissions from the same location but no enhanced emissions at OI1356~Å in the same observations was consistent with a signal from electron-impact dissociative excitation of H$_2$O, as measured in laboratory studies \citep{makarov04}. OI1304~Å and OI1356~Å emissions elsewhere on Europa during the December 2012 observations could be well explained by the global O$_2$ atmosphere \citep{roth14-science,Roth2021-Eur}. 

A direct follow-up campaign with HST/STIS to test a hypothesized connection of plume activity to periodic tidal stresses did not provide confirmation \citep{roth14-apocenter}. Although various studies later claimed further evidence for plumes at different locations on Europa \citep[e.g.,][]{sparks16,jia18,paganini2020}, there has been no confirmed detection or consistent picture of plume activity at Europa \citep{giono20,roth2025}. 

We present an analysis of 23 Ly$\alpha$ (1216~Å) images of Europa extracted from HST/STIS data taken in the spectral G140L setup in 1999 and on various dates between 2012 and 2020. The simultaneously observed oxygen emissions were published in \cite{roth16-eur} and \cite{Roth2021-Eur} for the datasets taken until 2015. The Ly$\alpha$ emissions from the first five datasets were included in the analyses in \citet{roth14-science} and \citet{roth14-apocenter} and are re-analyzed here. The 23 datasets presented were all taken when Europa was sunlit (not eclipsed) and not in transit of Jupiter. We focus on two different aspects in the Ly$\alpha$ signal: (1) Resonant scattering emissions from Europa's H exosphere, which are expected to be globally abundant and persistent; (2) Potential emissions from water vapor plumes, which would be localized and likely transient. 

%--------------------------------------------------------------------
\section{Observations and data analysis}
\label{sec:obs}

\begin{figure*}
   \centering
   \includegraphics[width=\textwidth]{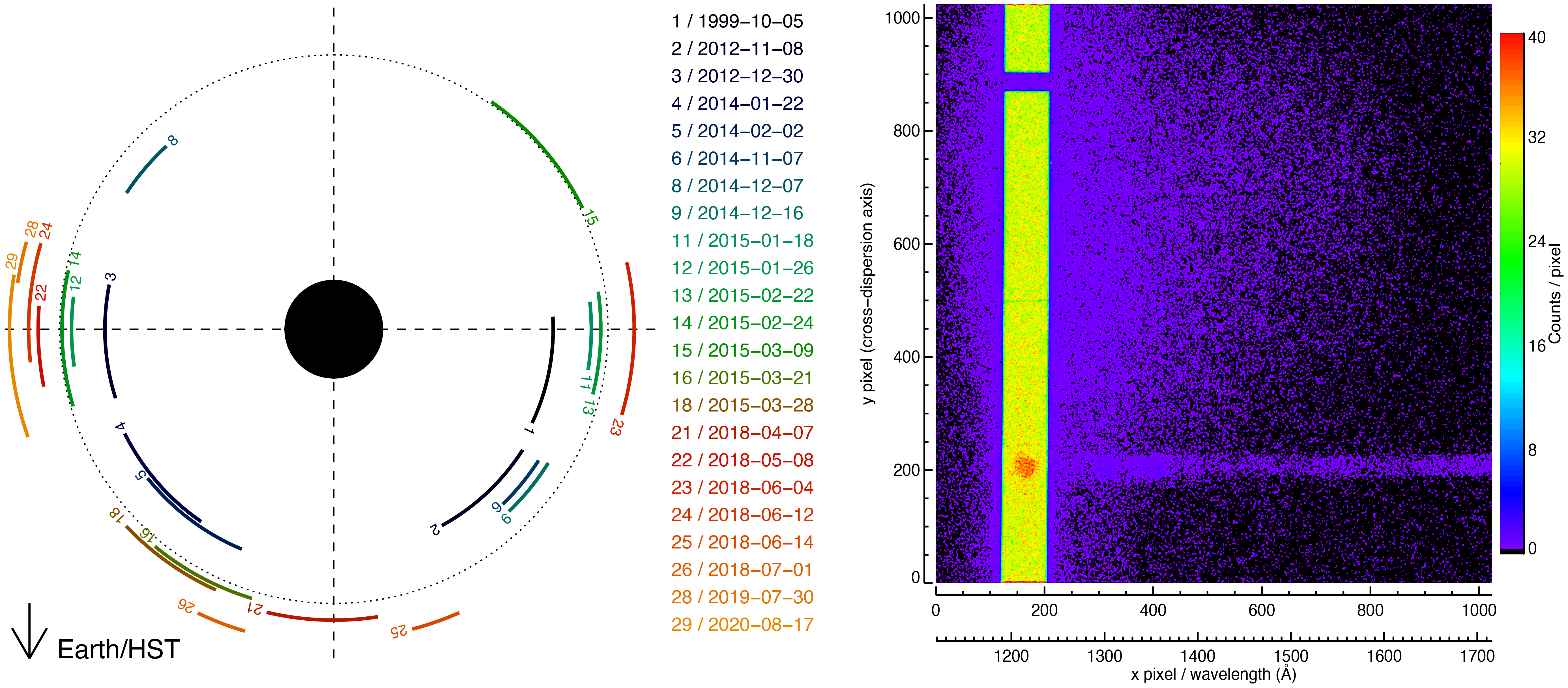}
   \caption{Overview of the analyzed HST/STIS observations: (Left) Orbital positions of Europa from the start of the first exposures to end of the last exposures for each of the 23 HST/STIS visits. Note that the gaps between exposures of a visit are not shown. Visit 21 combined exposures before and after transit. (Right) Complete STIS detector spectral image from visit 22. The Ly$\alpha$ signal from the geocorona and the interplanetary hydrogen fills the complete slit (yellow/green region). Reflected continuum sunlight from Europa's surface forms the trace around row y=200. Atmospheric oxygen emissions are present at 1304~Å and 1356~Å, exceeding the surface reflection.}
              \label{Fig:Orb}%
    \end{figure*}

We analyze all observations of Europa in sunlight (and not transiting Jupiter) taken by HST/STIS with the G140L grating and the 52"$\times$2" slit. Table \ref{tab:obs} and Figure \ref{Fig:Orb}(left) provide an overview of the observing parameters and geometries of the 23 analyzed datasets. During each visit between 2 and 10 exposures were taken by STIS with varying exposure time, within two to five HST orbits. In the visits after 2014, the capability of HST to observe moving targets over several hours (or several orbits of HST around Earth) was limited. As a result, most visits after 2014 are shorter with fewer exposures. The water vapor signal derived in \citet{roth14-science} was observed during visit 3 in December 2012 when Europa was near eastern elongation. Orbital positions similar to visit 3 were thus preferred in the follow-up HST campaigns, leading to a clustering of visits near maximum Eastern elongation (Figure \ref{Fig:Orb}). 

We combine all exposures taken during a visit into one observation to maximize the signal.
%The Ly$\alpha$ emissions might be varying between the exposures for a number of reasons. %The plasma environment is changing governed by the 13-h synodic period of the magnetospheric rotation affecting the atmosphere and emissions \citep{roth16-eur}. The sub-observer central meridian longitude changes by up two 30$^\circ$ from the first to the last exposure, related to possible changes due to inhomogeneous albedo or atmosphere distribution. 
Both the observing geometry and plasma environment change during a visit as indicated by the ranges in sub-observer W longitude and Jupiter's System-III longitude at Europa in Table \ref{tab:obs}. However, the gain in signal by combining the exposures is needed and possible variability is neglected as done in previous studies \citep{roth14-science}. Figure \ref{Fig:Orb}(right) shows an example detector image of the superposition of the three exposures taken during visit 22 in 2018. 
\begin{table*}
\caption{Parameters of the 23 analyzed HST/STIS observation visits.}
\label{tab:obs}
%\resizebox*{1.\textwidth}{!}{
\begin{tabular}{llccccccccc}
\hline
 &   & Start	&  End  & No. & Used & Distance & Europa   & Spatial    & Europa   & System-III\\
Visit$^{a}$ & Date & time	&  time & of  & exp.time	 & to Sun   & diameter & resolution & CML	    & longitude	\\ 
	  & (Start)	   & (UTC)	&  (UTC) & exp. &[min]		 & [AU]     & [arcsec] & [km/pixel] & [$^\circ$]& [$^\circ$] \\		
\hline
 1 & 1999-10-05 & 08:39 & 15:27 &  9 &    132.7 & 4.96 & 1.07 & 71.5 & 245 - 273 & 300 - 157 \\
 2 & 2012-11-08 & 20:41 & 03:28 & 10 &    155.0 & 5.04 & 1.04 & 73.9 & 209 - 238 &  24 - 242 \\
 3 & 2012-12-30 & 18:49 & 01:39 &  9 &    164.1 & 5.06 & 1.02 & 74.9 &  79-108 & 360 - 220 \\
 4 & 2014-01-22 & 14:02 & 20:53 & 10 &    143.4 & 5.20 & 1.01 & 76.0 & 117-146 & 201 - 61 \\
 5 & 2014-02-02 & 08:19 & 15:07 & 10 &    123.4 & 5.20 & 0.99 & 77.3 & 129-157 & 199 - 57 \\
 6 & 2014-11-07 & 16:02 & 19:14 &  3 &     30.7 & 5.30 & 0.81 & 94.7 & 224-237 &  68 - 171 \\
 8 & 2014-12-07 & 20:25 & 23:50 &  3 &     68.3 & 5.32 & 0.89 & 86.6 &  42-57 & 245 - 354 \\
 9 & 2014-12-16 & 17:46 & 21:09 &  3 &     65.8 & 5.31 & 0.91 & 84.3 & 224-238 & 244 - 352 \\
11 & 2015-01-18 & 00:59 & 04:36 &  3 &    102.7 & 5.33 & 0.98 & 78.6 & 261-276 & 250-6 \\
12 & 2015-01-26 & 22:17 & 01:55 &  3 &    107.0 & 5.33 & 0.99 & 77.9 &  83-98 & 244-1 \\
13 & 2015-02-22 & 11:00 & 16:17 &  4 &    154.6 & 5.34 & 0.98 & 78.2 & 256-278 & 132-301 \\
14 & 2015-02-24 & 05:58 & 12:51 &  5 &    195.3 & 5.34 & 0.98 & 78.5 &  77-107 &  67-288 \\
15 & 2015-03-09 & 01:03 & 08:01 &  5 &    193.4 & 5.34 & 0.96 & 80.1 & 296-325 & 189-52 \\
16 & 2015-03-21 & 16:59 & 22:21 &  4 &    145.0 & 5.34 & 0.93 & 82.4 & 141-163 & 209-21 \\
18 & 2015-03-28 & 17:42 & 22:59 &  4 &    124.0 & 5.34 & 0.91 & 84.0 & 133-156 & 216-25 \\
21 & 2018-04-07 & 11:00 & 16:15 &  2 &     70.0 & 5.41 & 0.94 & 81.3 & 167-189 &  13-181 \\
22 & 2018-05-08 & 14:00 & 17:45 &  3 &    125.1 & 5.41 & 0.98 & 78.5 &  85-101 & 195-314 \\
23 & 2018-06-04 & 01:36 & 08:34 &  4 &    162.0 & 5.40 & 0.96 & 80.1 & 253-283 &  45-268 \\
24 & 2018-06-12 & 22:34 & 03:57 &  3 &    109.6 & 5.40 & 0.94 & 81.4 &  74-96 &  29-201 \\
25 & 2018-06-14 & 03:11 & 05:23 &  2 &     73.5 & 5.40 & 0.94 & 81.5 & 195-204 & 226-296 \\
26 & 2018-07-01 & 11:37 & 13:48 &  2 &     75.2 & 5.39 & 0.90 & 84.9 & 154-164 & 250-321 \\
28 & 2019-07-30 & 04:21 & 06:09 &  2 &     48.7 & 5.28 & 0.94 & 82.0 &  74-82 & 309-7 \\
29 & 2020-08-17 & 02:26 & 09:21 &  4 &    125.4 & 5.14 & 1.00 & 76.6 &  80-110 &  48-268 \\
\hline
%\multicolumn{10}{l}{$^{a}$Footnote?.}
\end{tabular}
$^{a}$Visit numbers are identical to those used in \citet{roth16-eur} and \citet{Roth2021-Eur}. Visits taken when Europa was eclipsed by Jupiter (7, 10, 17, 19, 20, and 27) are not included here and thus missing in the otherwise consecutive numbering.
%}
\end{table*}

The Ly$\alpha$ geocorona, i.e., emissions from Earth's extended H exosphere, fill the entire slit (yellow in Figure \ref{Fig:Orb} right). The geocorona is particularly bright when HST is on the dayside of the Earth. As in previous studies \citep{roth14-apocenter}, we have therefore selected exposures with low geocorona signal or partially cut exposure durations to remove high geocorona periods. For this, the background brightness along the Ly$\alpha$ slit away from Europa was monitored using the STIS time-tag data and only periods where the measured background is lower than 10 kR (kiloRayleigh) were used. This threshold has been applied to all datasets except for visit 3. To allow direct comparison to the results in \citet{roth14-science} we use the full exposure time as used in that study. The background was however generally low during visit 3 except for the first 8 minutes in 3 out of the 9 exposures total (24 minutes high geocorona vs 140 minutes low geocorona). The total exposure time of the used data are given for each visit in Table \ref{tab:obs}. 

In addition to the geocoronal background, the slit centered on the Ly$\alpha$ wavelength contains contributions from interplanetary H, sunlight reflected from Europa’s surface, and Europa’s own H exosphere. Possible localized emissions may also be present near Europa, produced by electron excitation of H-bearing molecules -- such as the proposed water vapor aurora associated with a plume \citep{roth14-science}. Europa’s exact position on the detector is determined from the reflected sunlight signal, following the approach used in previous STIS studies. To disentangle these various Ly$\alpha$ components and accurately determine Europa’s detector position, we construct a model image (see Section \ref{sec:mod}) that includes all known Ly$\alpha$ sources except potential localized emissions (like plume aurora). The model image is also used to find the positions of Europa's disk on the detector. 

The model includes several fitting parameters that primarily scale the relative contributions of the different Ly$\alpha$ sources. In addition, various central ($x,y$) pixel positions for Europa’s disk are tested. Because Europa’s actual position on the STIS detector differs from that specified in the data file header, it must be determined directly from the observed signal. Accurate localization is particularly important when searching for potential localized emissions near the disk’s limb, where a steep gradient exists between on-disk and off-disk signals \citep{giono20}. To further constrain the disk center, the surface reflection at longer wavelengths ($\lambda$=1430–1530~Å), where no atmospheric emissions are present, is used to independently assess the central $y$ pixel.  The $x$ pixel cannot be constrained with the continuum trace, only at Ly$\alpha$. Given HST's pointing stability of $\sim$0.005 arcsec (or 1/5 of a STIS pixel), the position of the moon on the detector can be assumed stable over a visit.

%We focus on the observations in sunlight, where the scattering and the reflection of the sunlight is the dominant signal from Europa. Eclipse observations need to be processed and analyzed differently. Position finding is different. H corona should not be there, aurora might change if atmosphere changes in eclipse etc.

\section{Model for the Ly$\alpha$ emissions in the STIS images}
\label{sec:mod}

The constructed model consists for three components considering different sources of Ly$\alpha$ emission. The first component of the model represents the background and foreground emission present throughout the slit. This contribution is assumed to be constant along the $x$-direction (i.e., from the left to the right edge of the Ly$\alpha$ slit; yellow in Figure~\ref{Fig:Orb}, right panel). Along the slit ($y$-direction), the intensity is modeled with a second-order polynomial function to capture variations in brightness. This component includes geocoronal foreground emission as well as emissions from the hydrogen in the interplanetary medium (IPH) both in front of and behind Europa. The interplanetary medium is the interstellar gas cloud passing through Solar System and the hydrogen component scatters the solar Ly$\alpha$ flux \citep{izmodenov_distribution_2013}.

Because Europa’s disk blocks the background, a constant value -- representing the interplanetary H brightness derived from a model by \citet{Pryor2024} -- is later subtracted on the disk. The three coefficients of the polynomial along $y$ are fitted (independently) for each dataset.

The second model component is the signal from Europa's H exosphere. The measurements in attenuation against Jupiter \citep{roth17-europa} showed that the exospheric H number density $n_H$ is consistent with an escaping (comet-like) profile that falls off with $\frac{1}{\rho^2}$, where $\rho$ is the radial distance to the moon center. When integrating this number density profile along the line-of-sight from a distant observer (infinity) to the moon surface (for pixels on the disk where $r \leq R_E$), or to far beyond the target (to infinity, outside the disk $r > R_E$), the resulting line-of-sight H column density in the image plane around the moon becomes
   \begin{eqnarray}
     N_{\mathrm{H}}(r) = 
     \begin{cases}
        n_{\mathrm{H},0}\frac{R_E^2}{r}\arcsin\left({\frac{r}{R_E}}\right) = 
        N_{\mathrm{H},0}\frac{R_E}{r}\arcsin\left({\frac{r}{R_E}}\right) &  \text{if } r \leq R_E  \\
        n_{\mathrm{H},0}\frac{R_E^2}{r}\pi =
        N_{\mathrm{H},0}\frac{R_E}{r}\pi &  \text{if } r > R_E ,
    \end{cases}   
    \label{eq:Nexo}
   \end{eqnarray}
with the radial distance to Europa's center in the image plane $r$, Europa's radius $R_E = 1560.8$~km, the peak number density at the surface $n_{\mathrm{H},0}$, and the vertical column density 
\begin{equation}
N_{\mathrm{H},0} = n_{\mathrm{H},0}\,R_E  \: .        
\end{equation}
This profile (Eq. 1) assumes that the H atoms escape to infinity with constant velocity and thus neglects Europa’s (and Jupiter’s) gravity, as well as other processes such as radiation effects. Despite these simplifications, we adopt this profile for its simplicity and reproducibility.

Europa's H exosphere can further be assumed optically thin, with a line center optical depth of $\tau \approx 0.3$ at the line center for the vertical H column density and temperature derived here (Section \ref{sec:Hexo_col}) . We thus assume the same radial decrease for the intensity as a function of distance $r$ to the disk center in an image given as
  \begin{eqnarray}
     I_{\mathrm{H}}(r) = 
     \begin{cases}
        I_{\mathrm{H},0}\frac{R_E}{r}\arcsin\left({\frac{r}{R_E}}\right) & \text{ if } r \leq R_E\\
       I_{\mathrm{H},0}\frac{R_E}{r}\pi & \text{ if } r > R_E
    \end{cases}   
    \label{eq:Hexo}
   \end{eqnarray}
with the disk center brightness $I_{\mathrm{H},0}$.

For the third model component, which is the contribution from sunlight reflected off Europa’s surface, we follow the approach of previous studies \citep[e.g.,][]{roth14-science,Roth2021-Eur}. A high-resolution solar spectrum from the SOlar Stellar Irradiance Comparison Experiment (SOLSTICE) on the Solar Radiation and Climate Experiment (SORCE) \citep{mcclintock05} is mapped onto Europa’s disk, accounting for the spectral dispersion of the STIS configuration. Because the solar flux varies with solar longitude, we select a SOLSTICE spectrum from a nearby date when the solar longitude visible from Earth corresponded to that illuminating Europa at the time of the observation \citep{Roth2021}.

The surface reflection model incorporates both lateral albedo variations, due to changes in surface composition and texture, and the spherical geometry of Europa. Because the Ly$\alpha$ reflectivity is roughly anti-correlated with the visible albedo \citep{hendrix05,roth14-science,becker18}, we invert the USGS visible albedo map of Europa \citep{usgs_eurmap}, adjust the contrast, and project it onto the disk for the specific viewing geometry. Reflection from the rough, spherical surface is then simulated using the \citet{oren94} model, as applied to Europa in previous studies \citep{sparks16,giono20,Roth2021-Eur}.

By jointly comparing all datasets, we find the best agreement between the model and STIS images for a slightly reduced albedo contrast -- where the maximum reflection is about five times the minimum -- and a surface roughness parameter of 0.8. This roughness value is somewhat higher than the 0.57 adopted in previous works \citep{sparks16,Roth2021-Eur}, resulting in a disk that appears more uniformly bright, i.e., closer to the uniform disk assumed in earlier studies \citep[e.g.,][]{roth14-science}. The final reflected-sunlight model is scaled by the albedo, which is treated as a free parameter and fitted individually for each dataset. Although significant albedo variations between observations (especially those with similar geometry) are not expected, the albedo is fitted separately to ensure correct scaling of the reflected sunlight component in each case.

Model images are first generated with a 3$\times$ higher resolution than the STIS data resolution (the latter is roughly 40 pixels across Europa), primarily in order to ensure a smooth transition at the limb from on-disk to off-disk in the resulting model images. When all contributions are combined the spatial resolution is adjusted to the STIS data and convolved with the STIS point-spread-function at the Ly$\alpha$ wavelength \citep{krist11}. All model components with their parameters (exosphere brightness, surface albedo, background polynomial) are then fitted simultaneously for best agreement with the observed 2D STIS image within the slit. 

In the observations, Europa is located in the lower part of the slit (except in the 1999 data) to avoid regions on the detector affected by fiducial bars and dark noise (see Figure \ref{Fig:Orb}). For the fit, all pixels from the disk center row up to 200--300 pixels away from the disk center in both directions (and at least 5 pixels away from the detector edge) were included, with the chosen range depending on the disk position and noise level. 

   \begin{figure*}
   \centering
   \includegraphics[width=0.8\textwidth]{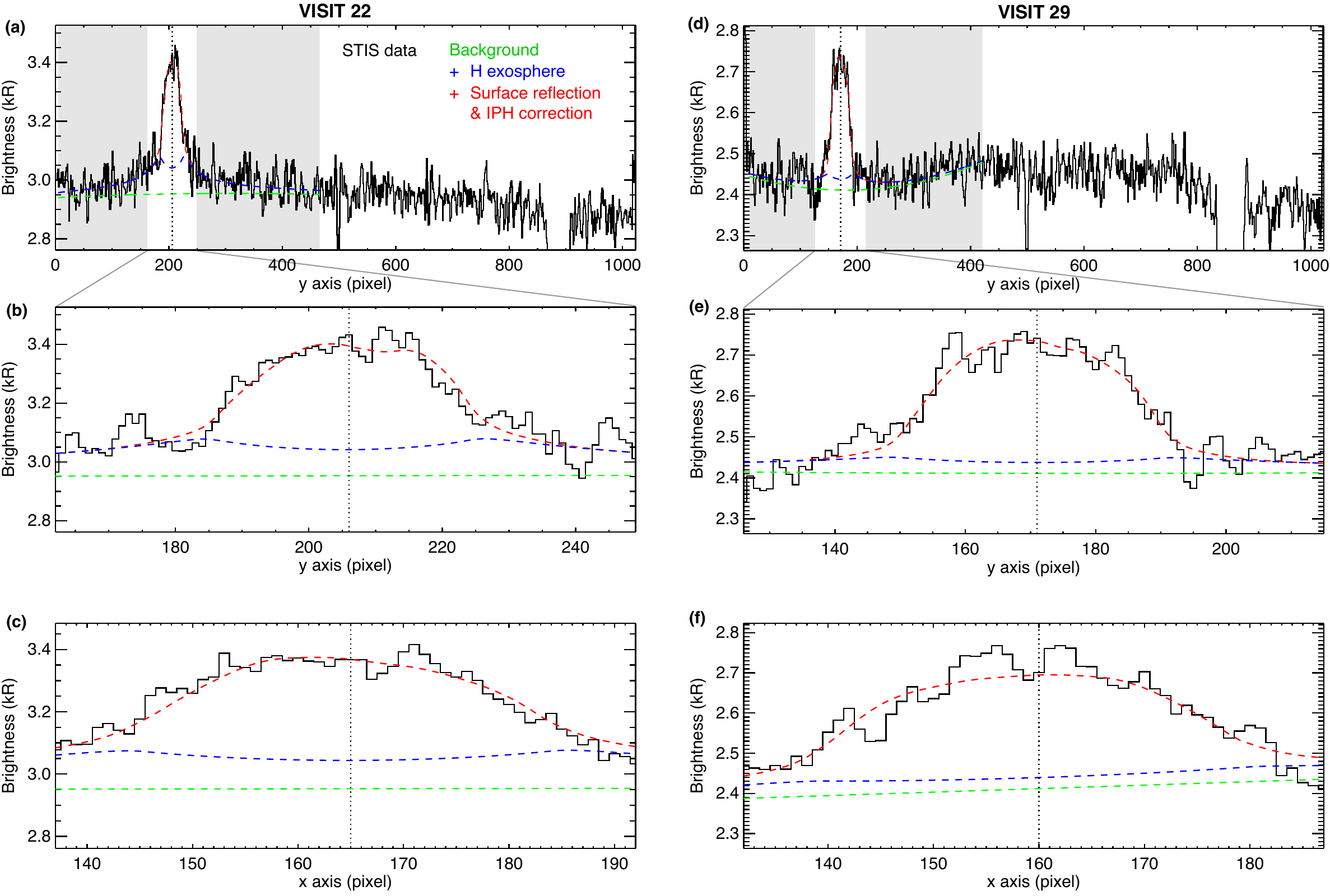}
   \caption{Data analysis and model fitting: (Top) Brightness for the entire Ly$\alpha$ slit along the $y$ axis, averaged over the $x$ axis (i.e., over the slit width). (Middle) Zoom-in of the top panel. (Bottom) Brightness along the $x$ axis, averaged over a $y$ range covering the disk diameter. The STIS data are shown as a black histogram. The fitted model is represented by dashed lines corresponding to different components: green for the background only, blue for the background plus H exosphere contribution, and red for the full model including surface reflection and IPH correction (see text). Note that the profiles are smoothed due to averaging over a detector axis (and thus across the disk), as well as by convolution of the model image with the STIS point-spread function. The vertical dotted lines indicate the disk center in all panels. The model was fitted within the central disk and the gray shaded regions (see top panel).}
    \label{Fig:lya_prof}% 
    \end{figure*}

Figure \ref{Fig:lya_prof} shows the data and the fitted model in averaged brightnesses in Rayleigh (1~R = $\frac{10^6}{4\pi}$~$\frac{\mathrm{photons}}{\mathrm{cm}^2\mathrm{s\,sr}}$) along the detector $y$ and $x$ axes for two example visits. There is a certain degeneracy when fitting the H exosphere profile and background simultaneously but we find very similar results for H density for different choices of $y$ ranges for the fit, which gives confidence in the robustness. 

We use the model to find a ($x,y$) position of Europa's disk center. The best position is assumed to be the one that minimizes the sum over all pixels (within 1.4~R$_E$ of the disk center) of the squared differences between observed and model pixel brightness. As an independent check, we find that the fitted surface albedo is also highest for the best-fit position in almost all datasets, as expected for the best alignment. For most images, the sum of the squared residuals varies only marginally in a range of 2 pixels around the minimum, suggesting an uncertainty of $\pm$2 pixels in $x$ as well as $y$ position. 

We then use the surface reflection continuum trace at longer wavelengths ($\lambda = 1430$–1530~Å; see Figure~\ref{Fig:Orb}, right) -- where no atmospheric emissions are expected to be present -- to provide an independent constraint on the $y$ position. For this purpose, we fit a linear function plus a Gaussian to the brightness profile along $y$, integrated over the wavelength range ($x$), where the center of the Gaussian is adopted as the best-fit $y$ position.

For most datasets, the $y$ position derived from the reflected continuum agrees with the Ly$\alpha$-based value to within one pixel. In cases where the two estimates differ by two to three pixels, we adjust the Ly$\alpha$ $y$ position by one pixel toward the continuum-derived value. For example, during visit 22, the Ly$\alpha$ fit suggested a central position at $y = 205$, while the continuum fit yielded $y = 207$; we therefore adopt $y = 206$ for the subsequent analysis, as shown in Figure~\ref{Fig:lya_prof}b.

\begin{table*}
\caption{Extracted brightnesses of Ly$\alpha$ sources, surface albedo values, and relative line-of-sight velocities to the Earth and Sun.}
\label{tab:res}
%\resizebox*{1.\textwidth}{!}{
\begin{tabular}{llccccccc}
\hline
Visit &  Date  & Foreground$^{a}$ & IPH background$^{b}$ & Surface reflection$^{c}$ & Albedo & H exosphere$^{d}$ & $v_{rad,geo}$ & $v_{rad,sun}$\\	
\# & YYYY-MM-DD & (R) & (R) & (R) & (\%) & (R) & (km/s) & (km/s) \\
\hline
 1 & 1999-10-05 & 4052 & 287 & 1364 & 1.69 ($\pm$0.02) & 105.7 ($\pm$10.5) &   4.7 &  13.8\\
 2 & 2012-11-08 & 3119 & 215 & 1226 & 1.84 ($\pm$0.02) & 46.4 ($\pm$9.8) &  -2.6 &  10.5 \\
 3 & 2012-12-30 & 4850 & 225 &  805 & 1.33 ($\pm$0.03) & 45.7 ($\pm$11.2) &   1.5 & -13.0 \\
 4 & 2014-01-22 & 4242 & 297 &  947 & 1.32 ($\pm$0.03) & 45.9 ($\pm$10.4) &  -0.9 & -10.4 \\
 5 & 2014-02-02 & 4861 & 316 &  998 & 1.48 ($\pm$0.03) & 109.1 ($\pm$13.0) &   6.4 &  -8.9 \\
 6 & 2014-11-07 & 7754 & 348 & 1264 & 2.19 ($\pm$0.12) & 139.0 ($\pm$33.6) & -16.7 &  12.3 \\
 8 & 2014-12-07 & 4051 & 384 &  878 & 1.47 ($\pm$0.05) & 131.5 ($\pm$15.8) & -35.3 & -11.2 \\
 9 & 2014-12-16 & 4621 & 382 & 1304 & 2.02 ($\pm$0.05) & 136.8 ($\pm$18.2) & -12.7 &  12.1 \\
11 & 2015-01-18 & 3185 & 401 & 1291 & 2.03 ($\pm$0.03) & 67.5 ($\pm$13.4) &   3.5 &  14.3 \\
12 & 2015-01-26 & 3179 & 343 &  959 & 1.70 ($\pm$0.04) & 137.3 ($\pm$10.6) & -19.3 & -13.1 \\
13 & 2015-02-22 & 3532 & 342 & 1184 & 2.06 ($\pm$0.03) & 115.5 ($\pm$10.1) &  22.5 &  14.2 \\
14 & 2015-02-24 & 3588 & 355 &  892 & 1.56 ($\pm$0.03) & 67.5 ($\pm$8.3) &  -4.0 & -13.1 \\
15 & 2015-03-09 & 4034 & 373 & 1163 & 1.94 ($\pm$0.03) & 137.1 ($\pm$9.4) &  26.4 &  12.1 \\
16 & 2015-03-21 & 3330 & 355 &  938 & 1.58 ($\pm$0.03) & 127.8 ($\pm$9.9) &  13.8 &  -7.8 \\
18 & 2015-03-28 & 4083 & 384 &  954 & 1.58 ($\pm$0.04) & 129.0 ($\pm$10.8) &  14.5 &  -9.3 \\
21 & 2018-04-07 & 2486 & 311 &  915 & 2.17 ($\pm$0.06) & 62.2 ($\pm$11.8) & -17.0 &   0.3 \\
22 & 2018-05-08 & 2639 & 314 &  740 & 1.74 ($\pm$0.04) & 57.0 ($\pm$9.0) & -14.8 & -14.2 \\
23 & 2018-06-04 & 3229 & 319 &  922 & 2.12 ($\pm$0.04) & 64.4 ($\pm$8.9) &  25.8 &  13.1 \\
24 & 2018-06-12 & 3343 & 319 &  681 & 1.60 ($\pm$0.05) & 0.0  ($\pm$12.9) &   2.1 & -13.9 \\
25 & 2018-06-14 & 3405 & 318 & 1031 & 2.41 ($\pm$0.06) & 69.2 ($\pm$13.1) &  20.6 &   2.3 \\
26 & 2018-07-01 & 2318 & 309 &  768 & 1.84 ($\pm$0.05) & 75.5 ($\pm$11.0) &  16.8 &  -7.4 \\
28 & 2019-07-30 & 3063 & 322 &  704 & 1.63 ($\pm$0.07) & 46.8 ($\pm$15.7) &   7.2 & -13.4 \\
29 & 2020-08-17 & 2087 & 324 &  668 & 1.52 ($\pm$0.03) & 17.3 ($\pm$8.3) &   1.7 & -14.4 \\
\hline
\end{tabular}
$^{a}$Foreground brightness is extracted from the data away from Europa. $^{b}$IPH brightness is modeled \citep{Pryor2024}. $^{c}$Surface reflection is the disk-average brightness of only the solar reflected component in the model. $^{d}$The H exosphere brightness is the disk center value corresponding to the vertical column density, see Eq. \ref{eq:Nexo}. 
\end{table*}

\section{Hydrogen exosphere analysis}

\subsection{H exosphere brightness}

\begin{figure*}
   \centering
   \includegraphics[width=0.90\textwidth]{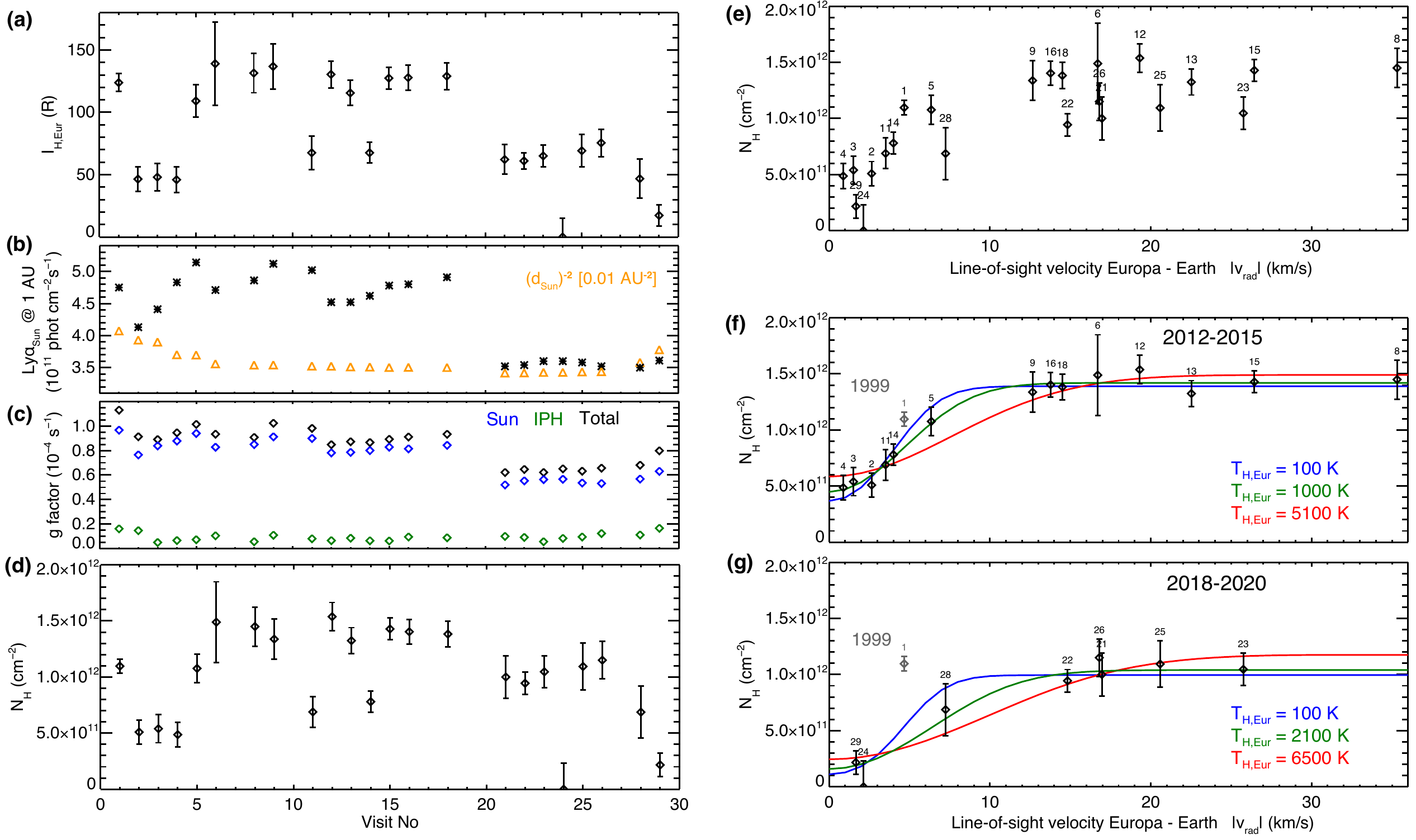}
   \caption{Analysis of Europa's H exosphere signal: (a) H exosphere brightness in the disk center from the fitted exosphere profiles (Eq. \ref{eq:Hexo}) for each visit. (b) Solar Ly$\alpha$ flux (black asterisks) measured at 1~AU near the observation date (see text). The inverse of the squared heliocentric distance of Europa (orange triangles, in unit 0.01~AU$^{-2}$ ) indicates the scaling of the flux used for calculating g-factors. (c) g-factors for resonant scattering of the sunlight (blue diamonds), of the IPH illumination (green) and combined (black). (d) H exosphere vertical column densities from conversion of measured brightnesses. (e) Same as in (d) but plotted against the line-of-sight velocity of Europa with respect to the Earth at the time of the observation. (f) and (g): same as (e) but only for visits 2-18 taken in 2012-2015 and for visits 21-29 taken in 2018-2020, respectively. Fitted attenuation profiles for different temperatures of the H in Europa's exosphere, see text for more information. The 1999 visit is shown in grey in (f) and (g) for comparison.}
              \label{Fig:H_exo}%
    \end{figure*}

The exosphere brightness peaks at the limb, where it is $I_{\mathrm{H}}(r=R_E) = I_{\mathrm{H},0} \, \pi$, and decreases to $I_{\mathrm{H},0}$ in the disk center ($r = 0$) (see Equation \ref{eq:Hexo}). For further analysis and interpretation, we use the value for the brightness at disk center $I_{\mathrm{H}}(r=0) = I_{\mathrm{H},0}$ from the fitted exosphere brightness profile as reference. Figure \ref{Fig:H_exo}a shows the retrieved disk center H exosphere brightness, $I_{\mathrm{H},0}$, for all 23 visits, ranging between 18~R (visit 29) and 139~R (visit 6). Notably, a zero brightness was fitted for visit 24.

\subsection{Conversion from brightness to column density}
\label{sec:Hexo_col}
We assume the H exosphere to be optically thin at the Ly$\alpha$ line and the observed emission to originate from a single resonant scattering process. The resonantly scattered sunlight from the H exosphere measured by STIS is then given by the photon scattering coefficient (or g-factor, $g_{Ly\alpha}$) times the column density
\begin{equation}
\label{eq:I_g_N}
I_{\mathrm{H}}(r) = g_{Ly\alpha}\,N_H(r) \quad , 
\end{equation}
connecting Equations \ref{eq:Nexo} and \ref{eq:Hexo}. The exosphere model part is scaled linearly with one parameter in the individual fit for each dataset, represented by the vertical H column density $N_{\mathrm{H},0}$ in Equation \ref{eq:Nexo}.

The strongest illumination source is the incoming solar flux, but we find that the illumination by the IPH can be an additional source and therefore consider it for the analysis. 
%Figure \ref{Fig:SunIPMGeo}a shows typical spectral fluxes from the Sun  \citep{Machol2019} and the IPM \citep{Pryor2024} at Jupiter. 
Thus, we estimate the scatterable spectral flux and the resulting scattering factors for the two sources, $g_{sun}$ and $g_{IPH}$. From these we calculate a combined resonant scattering factor as 
\begin{equation}
     g_{Ly\alpha} = g_{sun} +g_{IPH} = \left(\pi F_{sun} + \pi F_{IPH} \right)\frac{\pi e^2}{m_e c}f\frac{\lambda_0^2}{c},
\end{equation}
where $\pi F_{sun}$ and $\pi F_{IPH}$ values are the spectral fluxes from the Sun and IPH in the rest frame of Europa, $e$ is the electron charge, $m_e$ is the mass of the electron, $c$ is the speed of light, $f$ is the oscillator strength (0.416), and $\lambda_0$ is the Ly$\alpha$ wavelength (1216~Å). 
   \begin{figure}
   \includegraphics[width=0.5\textwidth]{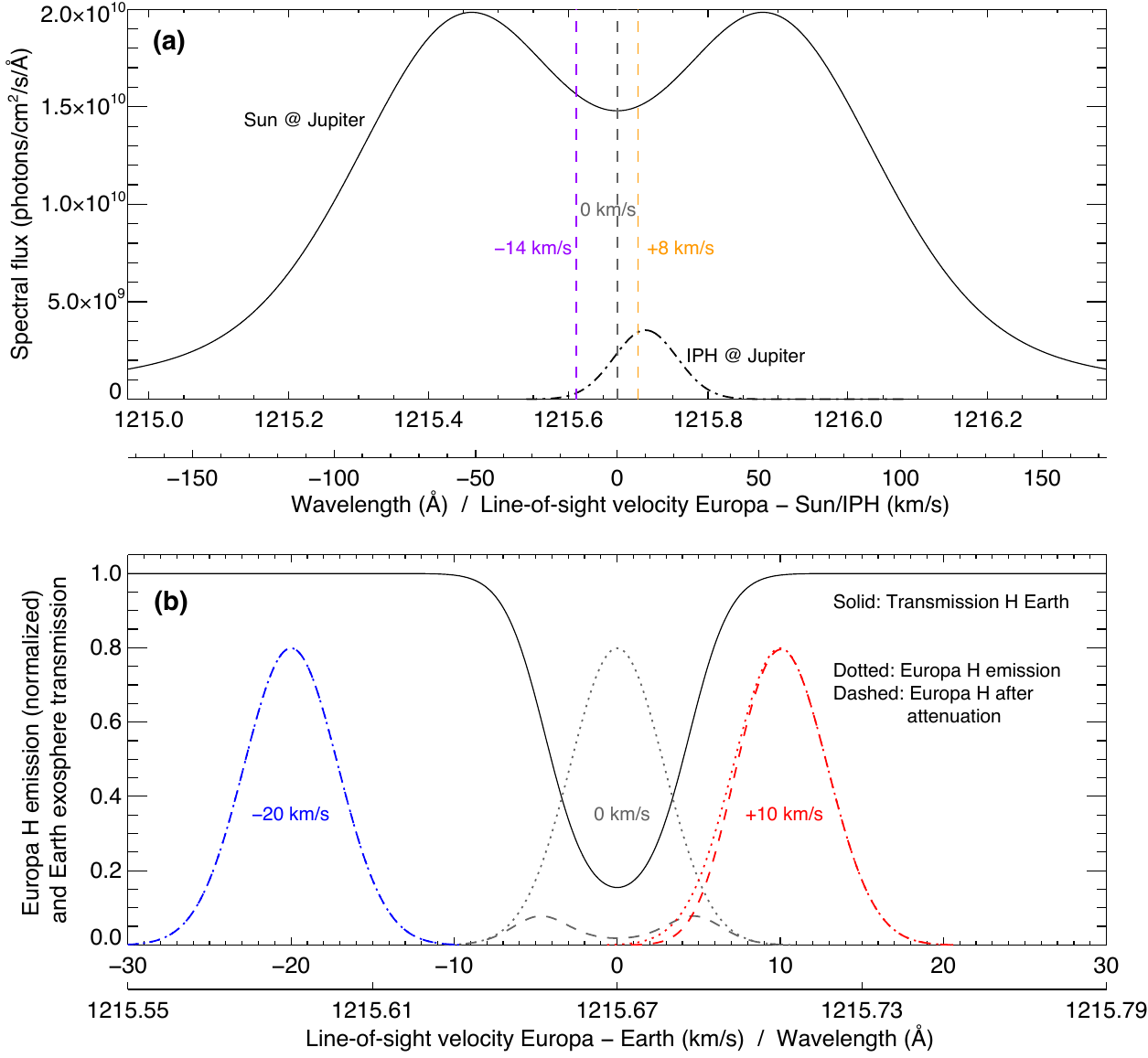}
   \caption{Analysis of the spectral aspects of the Ly$\alpha$ signals: (a) Typical Ly$\alpha$ spectral fluxes from the Sun (solid) and from the IPH (dash-dotted) at the heliocentric distance of Jupiter. The IPH flux is shown as a Gaussian profile with $T$ = 15000 K, at an offset to the solar line because of the IPH movement relative to the Sun. The scatterable flux by the H in Europa's exosphere varies with the relative line-of-sight velocity between the moon and the source (Sun, or IPH in a certain direction as shown for three example velocities). The vertical dashed lines (purple, grey, orange) show how the contributions from the Sun and the IPH can vary due to the respective Doppler shift. (b) Ly$\alpha$ transmission of the Earth H exosphere in the rest frame of Earth for an assumed column density of $1\times10^{13}$~cm$^{-2}$ and temperature of 1000~K (solid). Ly$\alpha$ emission profiles from Europa's H exosphere (T = 1000~K) from resonant scattering before (dotted) and after (dashed) attenuation in the Earth exosphere from the shown transmission. Three example profiles (blue, grey, red) are shown for three different relative line-of-sight velocities of Europa with respect to Earth. The strongly shifted emission (blue) is not affected by the Earth at all.}
    \label{Fig:SunIPMGeo}%
    \end{figure}

For the solar source, we use the line-integrated composite Ly$\alpha$ photon flux at 1~AU \citep{Machol2019} from the same date as for the surface reflection (considering the solar longitude variations), see Figure \ref{Fig:H_exo}b. The line-integrated flux is converted to spectral flux at the line center using the expression by \cite{Kretzschmar2018} (their equation 4). We then consider a small offset from the line center due to the radial velocity component of Europa with respect to the Sun, using a Lorentzian-Gaussian solar line shape as used by \cite{gladstone15}. Figure \ref{Fig:SunIPMGeo}a shows three example relative velocities and the corresponding changes in the solar spectral flux. The spectral flux at the Doppler shifted wavelengths is slightly higher than the line center spectral flux by up to 6\% (Figure \ref{Fig:SunIPMGeo}a). Finally, the flux is adjusted to the Sun-Europa distance (Figure \ref{Fig:H_exo}b and Table \ref{tab:obs}). 

%For the further analysis of the H exosphere signal, we use the fitted H exosphere emission profile for the selected best disk position. The emission profile is later converted to a density profile assuming 

For the IPH, we simulate the all-sky Ly$\alpha$ brightness using the model of \cite{Pryor2024}. The total flux from the dayside sky hemisphere (as seen from the sub-solar point on the surface) is typically near $4\times10^8$~photons/cm$^2$/s and hence about 2\% of the total solar flux. We assume a simple Gaussian line profile and a temperature of 15000~K for the IPH \citep{izmodenov_distribution_2013}. The narrower IPH spectral line compared to the solar line leads to a relatively higher line center spectral flux of $3.5\times10^9$~photons/cm$^2$/s/Å or $\sim$23\% of the solar line center spectral flux, see the maximum of the IPH curve and the value at the dip in the solar line center in Figure \ref{Fig:SunIPMGeo}a. The interplanetary hydrogen moves with a velocity of 20-25 km/s in the direction of ecliptic longitude and latitude of 72.5$^\circ$ and -8.9$^\circ$ (and the IPH profile is thus shown at a slight offset in Figure \ref{Fig:SunIPMGeo}a). 

To estimate the Doppler shift of the IPH hydrogen spectral line resulting from Europa’s relative motion, the radial velocity across each point in the sky must be considered \citep{joshi2025}. For each visit, we generate a sky map of the IPH brightness using the model of \citet{Pryor2024} (see Figure \ref{Fig:IPMmap}) and compute the relative velocity for each sky direction at the time of observation to estimate the IPH line flux that can be scattered by Europa’s H exosphere. When the brightest region of IPH illumination in the sky (near the Sun) is strongly Doppler-shifted, the scatterable IPH flux is low, as seen during visit 3. Conversely, when the IPH motion is largely perpendicular to the Sun–Europa line, the Doppler shift in the brightest region is minimal, and the scatterable IPH flux is high, as in visit 29.

Figure \ref{Fig:H_exo}c shows 
%the scatterable solar and IPM spectral fluxes for all datasets in panel b,  
the resulting g-factors for both the Sun and IPH for each visit. The H column densities after conversion with the g-factors (Eq. \ref{eq:I_g_N}) are shown in Figure \ref{Fig:H_exo}d. 

\subsection{Extinction in Earth exosphere and signal measured by HST }
\label{ssec:Hextinct}
Extinction of emissions from planetary H exospheres can occur in Earth's extended exosphere via scattering by atomic hydrogen \citep{alday17}. For simplicity and following \cite{alday17}, we assume that both the Europa exosphere emission profile and the profile of the attenuation by H at Earth are only Doppler-broadened due to the thermal motion of the H atoms. We further assume that the velocity profiles are described by Gaussian normal distributions. To convert the velocity profiles to spectral Ly$\alpha$ emission profile as a function of wavelength (or vice versa), the relative velocity $\Delta v$ is translated to wavelength shift as $\Delta\lambda = \frac{\Delta v}{c}\lambda_{Ly\alpha}$, with the speed of light $c$ and the Ly$\alpha$ vacuum wavelength $\lambda_{Ly\alpha} =$1215.667~Å. The velocity profile for the spectral brightness emitted from Europa's exosphere is then given by 
\begin{align}
\tilde I_{\mathrm{H}}(v) = \frac{I_{\mathrm{H},0}}{2\pi k_B T_{\mathrm{H},Eur} / m_H} \, \exp{\left(-\frac{(v-v_{0})^2}{2k_BT_{\mathrm{H},Eur}/m_H}\right)}  \\= 
\frac{g_{Ly\alpha}N_{\mathrm{H},Eur}}{2\pi k_B T_{\mathrm{H},Eur} / m_H} \, \exp{\left(-\frac{(v-v_{0})^2}{2k_BT_{\mathrm{H},Eur}/m_H}\right)} \, ,
\label{eq:emit}
\end{align}
with the reference velocity $v_0$, the H atomic mass $m_H$, temperature of Europa's H exosphere $T_{\mathrm{H},Eur}$, the Boltzmann constant $k_B$, and the line-integrated brightness $I_{\mathrm{H},0} = g_{Ly\alpha}N_{\mathrm{H},Eur}$. We note that Europa’s atmosphere is a rarified gas environment and thus not necessarily distributed thermally, but the emission Doppler width can still inform about an effective temperature \citep[e.g.,][]{Lierle2022}. Figure \ref{Fig:SunIPMGeo}b shows three examples of emission profiles (normalized to a maximum of 0.8 for visibility) for $T_{\mathrm{H},Eur} = 1000$~K and $v_0$ values of 0~km/s, +10~km/s and $-20$~km/s. 

The intensity measured by HST after extinction in the Earth exosphere, $I_{HST}$, is given by the Beer–Lambert law $I_{HST}(v) = I_{\mathrm{H}}(v) \exp\left( - \tau(v)\right)$. The optical depth $\tau$ is the product of the line-of-sight Earth exosphere column density, $N_{\mathrm{H},Geo}$ and the velocity dependent attenuation cross section 
\begin{equation}
\sigma_{scat}(v) = \sigma_{0} \, \exp{\left(-\frac{v^2}{2k_BT_{\mathrm{H},Geo}/m_H}\right)} \, ,
\end{equation}
with the line-center scattering cross section $\sigma_{0}$. We assume here the rest frame of Earth. The transmission for an assumed example column density of $N_{\mathrm{H},Geo}= 10^{13}$~cm$^{-2}$ and temperature $T_{\mathrm{H},Geo}=1000$~K of the Earth exosphere is shown by the solid black in Figure \ref{Fig:SunIPMGeo}b. In the Earth rest frame, the velocity offset of the Europa emission profile ($v_0$ in Eq. \ref{eq:emit}) is given by the line-of-sight component of the velocity of Europa with respect to Earth $v_{rad}$. The attenuated spectral brightness from Europa as measured by HST is then given by
\begin{align}
\label{eq:Hfinal}
\tilde I_{HST}(v) =  \frac{g_{Ly\alpha}N_{\mathrm{H},Eur}}{2\pi k_B T_{\mathrm{H},Eur} / m_H} \, \exp{\left(-\frac{(v-v_{rad})^2}{2k_BT_{\mathrm{H},Eur}/m_H}\right)} \\
\times \exp\left[ - N_{\mathrm{H},Geo} \sigma_{0} \, \exp{\left(-\frac{v^2}{2k_BT_{\mathrm{H},Geo}/m_H}\right)}\right]    \quad.
\end{align} 
The total measured brightness $I_{HST}$ for a given $v_{rad}$ is given by integrating the emission profile over the velocity. The total intensity of the example emission profiles in Figure \ref{Fig:SunIPMGeo}b are 31\% (gray, no offset), 95\% (red, 10~km/s redshift) and $>$99.9\% (blue, 20~km/s blueshift) of the original intensity, still assuming $N_{\mathrm{H},Geo}= 10^{13}$~cm$^{-2}$ and $T_{\mathrm{H},Geo}=1000$~K (solid line, Figure \ref{Fig:SunIPMGeo}b) 

For the analysis of the Europa exosphere intensity measured in the HST datasets, we fit the total measured intensity (i.e., Equation \ref{eq:Hfinal} integrated over $v$) to all datasets assuming the corresponding $v_{rad}$ at the time of the observations and $T_{\mathrm{H},Geo}=1000$~K. The fitted parameters are the vertical column density $N_{\mathrm{H},Eur}$ and temperature $T_{\mathrm{H},Eur}$ of Europa's exosphere as well as the Earth exosphere column density $N_{\mathrm{H},Geo}$. 

\subsection{Constraints on Europa's H exosphere}
\label{sec:Hconstraints}

A signal from Europa's H exosphere is derived (i.e., $I_{\mathrm H} > 0 $) in all visits, with the mentioned exception of visit 24 where the best fit of the model suggests that a measurable H exosphere signal is not present ($I_{\mathrm H} = 0$), see Figure \ref{Fig:H_exo}a. The highest brightnesses and thus vertical H column densities (Figure \ref{Fig:H_exo}d -- before considering extinction in the Earth exosphere) of around $1.5 \times 10^{12}$~cm$^{-2}$ were derived for visits 6 and 12.

When plotted against the (absolute) radial velocity of Europa with respect to Earth at the time of the observation, it is apparent that the measured signal is systematically lower at low Doppler shifts (Figure \ref{Fig:H_exo}e). In addition, the values from visits 21--29 (years 2018-2020) appear to be systematically lower than values from earlier visits (2012-2015 and 1999) at similar radial velocities. This difference was even more pronounced in the signal brightnesses (Figure \ref{Fig:H_exo}a) but has been balanced somewhat by the g factors, which were -- like the exosphere brightness -- systematically lower in the later years (2018-2020, visits 21-29, Figure \ref{Fig:H_exo}c) due to the low solar flux (Figure \ref{Fig:H_exo}b). The solar flux at Europa was particularly low in 2018 (visits 21 to 26), when the solar activity was near its minimum and Jupiter was near aphelion, i.e., farthest away from the Sun (Figure \ref{Fig:H_exo}b). 

In the following analysis, we analyze two periods (visits 2-18 in years 2012-2015, and visits 21-29 in years 2018-2020) separately, fitting the total intensity profiles (Section \ref{ssec:Hextinct}) individually for each period. Table \ref{tab:extinct_fit} shows the best fit values and the values for the maximum and minimum temperature $T_{\mathrm{H},Eur}$ considered to provide a resulting curve in reasonable agreement with the observations. The three profiles are shown in green (best-fit $T_{\mathrm{H},Eur}$), blue (minimum $T_{\mathrm{H},Eur}$), and red (maximum $T_{\mathrm{H},Eur}$) in Figure \ref{Fig:H_exo}f and g. 

\begin{table}[ht]
   \caption{Fitted parameters for column density and temperature of Europa's H exosphere, and for column density of Earth's H exosphere.}
     \label{tab:extinct_fit}
    \centering
    \begin{tabular}{lcccc}
    \hline
    Visits $\quad \quad$ &       $T_{\mathrm{H},Eur}$   & $N_{\mathrm{H},Eur}$ & $N_{\mathrm{H},Geo}$  &  $\chi^2_\nu$  \\
         
    2--18   &  1000   & $1.4 (\pm0.1) \times 10^{12}$  & $9.9 \times 10^{12}$ & 0.27 \\
\\ 
\multicolumn{1}{r}{$T_{min}$:} &   100   & $1.4 \times 10^{12}$  & $7.5 \times 10^{12}$ &  0.53\\          
\multicolumn{1}{r}{$T_{max}$:} &  5100   & $1.5 \times 10^{12}$  & $2.7 \times 10^{13}$ &  1.00 \\
   \hline
    21-29   &   2100   & $1.1(\pm0.1) \times 10^{12}$  & $3.9 \times 10^{13}$ & 0.56 \\ 
    \\
\multicolumn{1}{r}{$T_{min}$:}  &   100    & $1.0 \times 10^{12}$  & $1.2 \times 10^{13}$ & 0.81 \\   
\multicolumn{1}{r}{$T_{max}$:}  &   6500   & $1.2 \times 10^{12}$  & $5.6 \times 10^{14}$ & 1.00 \\
    \hline
    \end{tabular}
\end{table}

The fitted H column density for Europa is primarily constrained by the data points that are not affected by the attenuation. Thus, it is hardly correlated with the other fit parameters and well constrained. The resulting $N_{\mathrm{H},Eur}$ values differ between the two periods beyond the derived uncertainties (Table \ref{tab:extinct_fit}). 

Europa's exospheric H temperature is coupled to some extent to the column density of the Earth H exosphere: For higher Europa exosphere temperature and thus a broader emission line, a higher $N_{\mathrm{H},Geo}$ is needed to fit the maximum extinction observed at lower Doppler shifts. Furthermore, changes in $T_{\mathrm{H},Eur}$ most of all define how gradual (or steep) the transition from maximum attenuation to the unattenuated maximum signal is. This transition happens at Doppler shifts between $\sim$3~km/s and $\sim$10~km/s, where only few observations were taken (Figure \ref{Fig:H_exo}e, 2012-2015). We find that even a very low temperature of $T_{\mathrm{H},Eur} = 100$~K, as for an exosphere that is in thermal equilibrium with Europa’s surface, yields reasonable fits to the data. Maximum temperatures that yield $\chi^2$ values near 1.0 are 5100~K and 6500~K, respectively, for periods 1 and 2.  The difference in the resulting best-fit $T_{\mathrm{H},Eur}$ values between the periods (1000~K and 2100~K) is thus well within the uncertainties of the fitted temperature. 

\section{Search for localized (aurora) emissions}
\subsection{Global residual images}
The model generated as described in Section \ref{sec:mod} includes all known Ly$\alpha$ sources from Europa as well as from the background and foreground. For the search for localized emissions from, e.g., patchy aurora (from outgassing at plumes or other sources), we therefore subtract the modeled images from the STIS images and analyze the residual emissions.

\begin{figure*}
   \centering
   \includegraphics[width=0.9\textwidth]{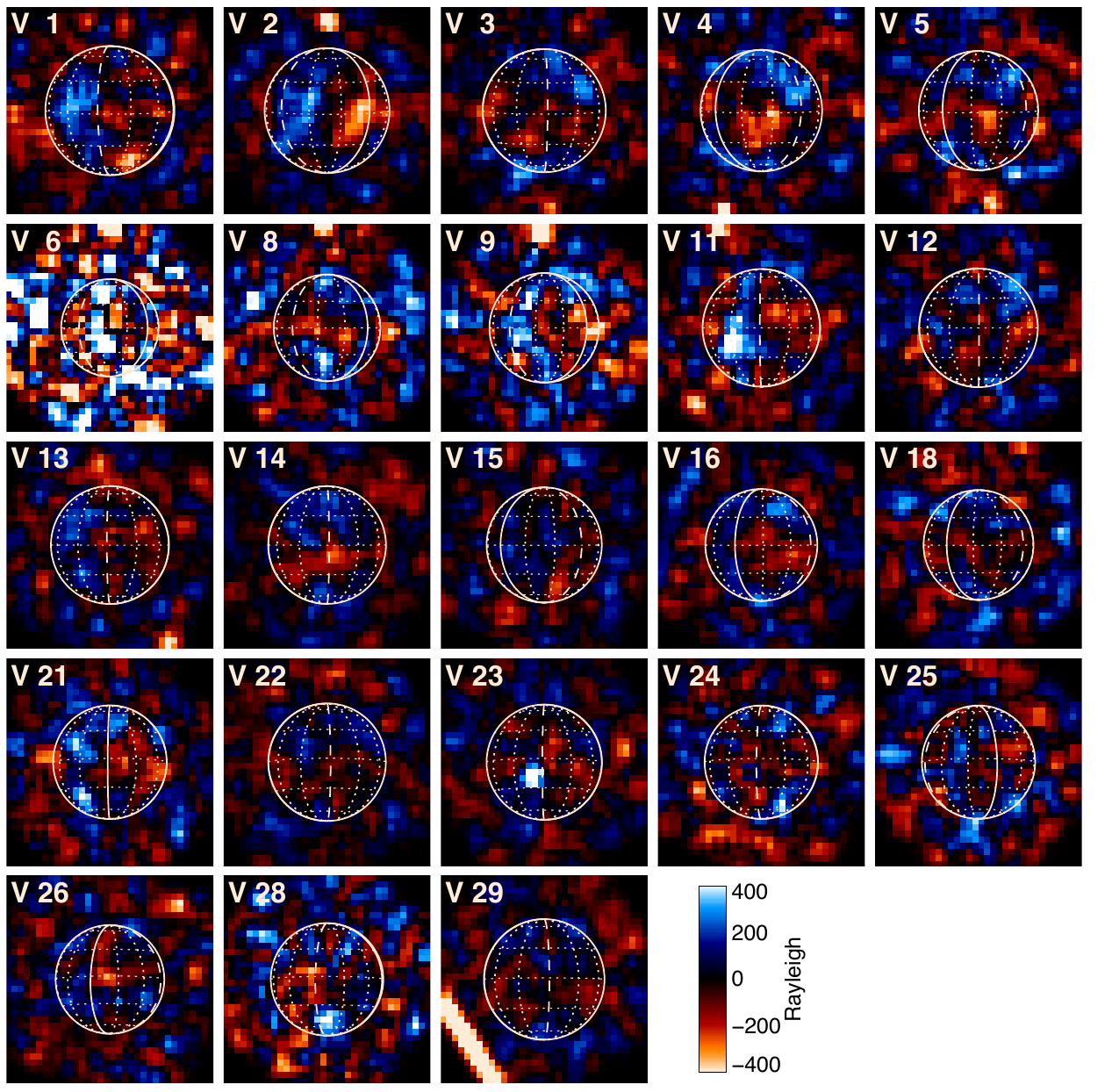}
   \caption{Residual Ly$\alpha$ images after subtraction of the model image. The images were binned (2x2 detector pixels) and boxcar smoothed (3x3 binned pixels) for display. The rotation from the detector to the Europa frame leads to missing signal in the corners, as seen in, e.g., the noisy image of visit 6 (V6).}
    \label{Fig:ResLyaImgs}%
    \end{figure*}

The modeled Ly$\alpha$ contributions are all relatively homogeneous, with only the reflected-sunlight component exhibiting structure due to the inverted albedo map used. Small-scale surpluses over regions smaller than Europa’s disk should therefore not appear in the model images -- particularly outside the disk -- and thus remain as surpluses in the residual-emission images. After subtracting the model, we extract a 72×72-pixel image centered on Europa for each visit and rotate the images so that the north pole points upward. All images are shown in Figure \ref{Fig:ResLyaImgs} using the same brightness color range from $-$400~R to $+$400~R. Because the subtracted model was fitted to match the global brightnesses, the residual emissions are distributed around zero.

The differences in appearance of the emissions originate primarily from the different noise levels, mostly determined by the level of the foreground emission (Table \ref{tab:res}) and the total used exposure time (Table \ref{tab:obs}). Higher foreground emission (from Earth exosphere) and shorter exposure times yield higher noise levels. Visit 6 exhibits the highest noise level because the observations were taken long before Jupiter opposition, when the geocorona emissions were high and substantial exposure cuts led to a short usable exposure time.    

The signal on Europa’s disk contains a substantial contribution from sunlight reflected off the icy surface, with intensities ranging from $\sim$0.7~kR on the leading side to roughly $\sim$1.3~kR on the trailing side (Table \ref{tab:res}). Any apparent surplus on the disk may therefore originate from, or at least be influenced by, local variations in the Ly$\alpha$ albedo that are not captured in the smoothed inverted visible maps used in the model. In some images, it appears that our model does not fully capture the spatial variation in surface reflection, leading to systematically brighter or darker residual regions on the disk. For example, in the image from visit 2 (Figure \ref{Fig:ResLyaImgs}, V2), the left (dawn) hemisphere is brighter and the right (dusk) hemisphere is fainter; similar patterns are seen in later trailing-side images, such as visit 11. Consequently, as in our previous studies \citep{roth14-science,roth14-apocenter}, we refrain from interpreting inhomogeneities in the residual Ly$\alpha$ brightness on the disk.

The most notable on-disk surplus emissions appear in visits 11 and 23, near the central longitude slightly below the equator. In visit 11, are larger region on the left part of the disk shows positive signal while the right side reveals negative residual signal. This indicates a misfit of the reflection model. During visit 23, several high-count pixels are randomly distributed across the detector, indicating that this feature is likely an artifact. Moreover, no co-located surplus is seen in the oxygen 1304 Å emission, which can serve as supporting evidence for Ly$\alpha$ enhancements produced by electron-impact dissociative excitation of H$_2$O \citep{roth14-science,Roth2021-Eur}.

\subsection{Limb bin analysis for all visits}
\label{sec:limbbin}
For a quantitative assessment of positive Ly$\alpha$ emission outliers, we again follow the previous studies and focus on the region above Europa’s limb between 1.0~$R_E$ and 1.25~$R_E$. As in \cite{roth14-science}, we subdivide this annulus into 18 azimuthal bins, each with an angular width of 20$^\circ$, and calculate the average residual brightness in each bin.
%Figure XX shows the Ly$\alpha$ images with the limb bins and the average brightness per bin with the propagated uncertainty for the two example visits 13 and 22.

\begin{figure}
   \centering
   \includegraphics[width=0.49\textwidth]{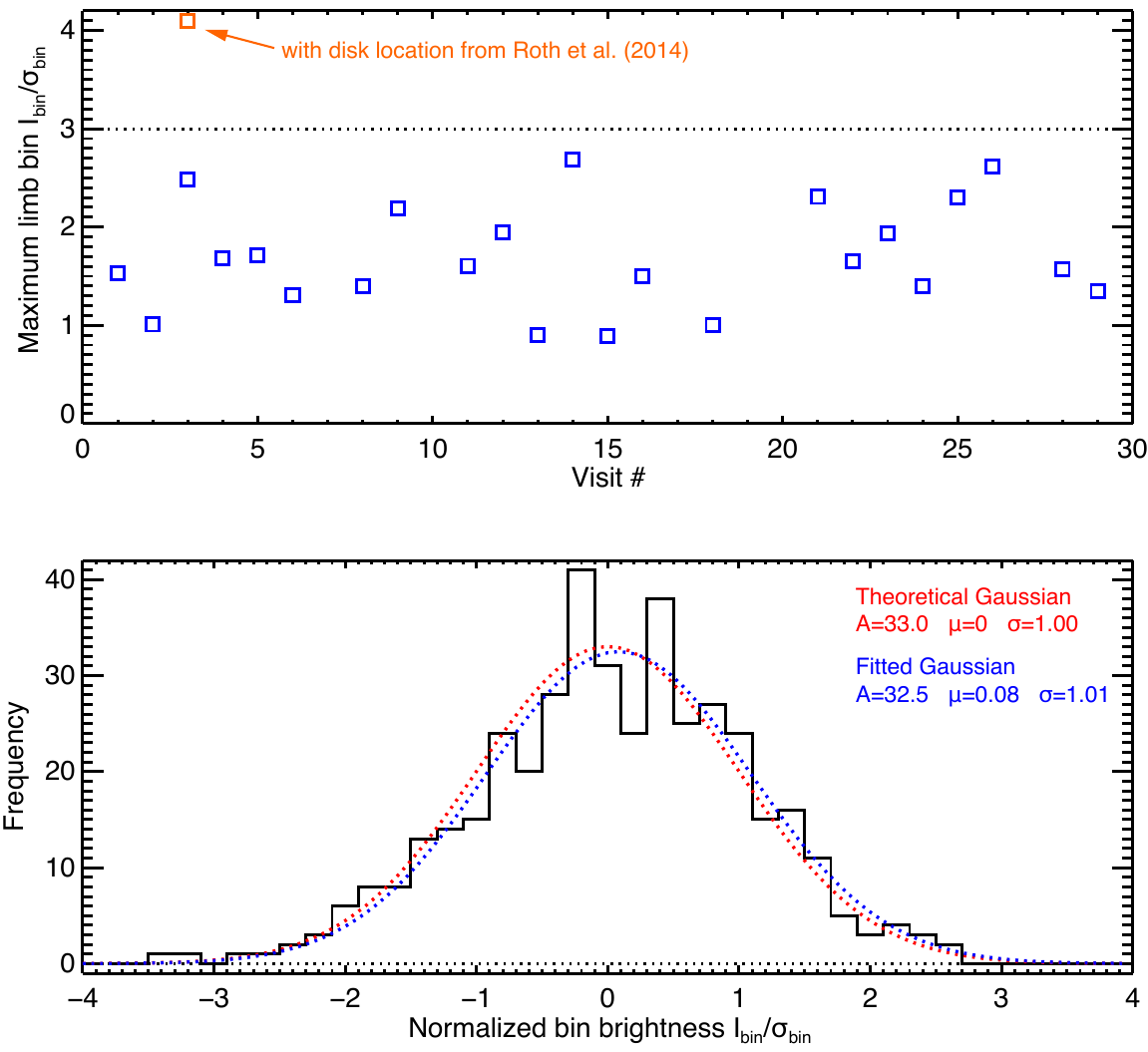}
   \caption{Statistical analysis of the limb bins: (Top) Brightness significance $I_{bin}/\sigma_{bin}$ of the limb bin with the highest value for each visit (blue squares). For visit 3, the orange square shows the value when assuming the disk position from \cite{roth14-science}. (Bottom) Histogram of the frequency of limb bin brightnesses for all 414 bins from all visits. The red dotted line shows the theoretical Gaussian distribution for sample size N=414. The blue dotted lines shows a Gaussian distribution fitted to the histogram. The values for max amplitude $A$, mean $\mu$, and standard deviation $\sigma$ are given for both Gaussians.}
    \label{Fig:limb_stats}%
    \end{figure}

In \cite{roth14-science}, the detection of a significant emission surplus was defined when the brightness in one bin exceeds the average of the other bins in the same image by more than three times the propagated statistical uncertainty ($>$3$\sigma$ detection). Figure \ref{Fig:limb_stats}a shows the bin brightness $I_{bin}$ divided by its statistical uncertainty $\sigma_{bin}$ for the brightest limb bin in each image. In contrast to the previous studies, we did not subtract the average brightness of the other bins of the same image in our analysis here, because this average is already very close to zero after considering and subtracting the H exosphere signal. 

In our analysis, the maximum value does not exceed the threshold of 3 in any of the visits, indicating that no statistically significant local surplus in Ly$\alpha$ emission is detected. This includes the image from visit 3 in December 2012, which was interpreted in our earlier study as showing an emission surplus. We further discuss the results for visit 3 and compare them to those in \cite{roth14-science} in Section~\ref{sec:vis3}. First, we investigate the statistical behavior of the limb-bin brightness to validate our results.

With 23 visits (or 23 images) and 18 limb bins in each image, we have a total of $N$=414 limb bin measurements. Figure \ref{Fig:limb_stats}b shows the frequency distribution of the obtained brightnesses significance values $I_{bin}/\sigma_{bin}$ in a histogram with a column width of $\Delta I_{bin}/\sigma_{bin}$=0.2. The distribution is in very good agreement with the theoretical Gaussian distribution of the form 
\begin{equation}
    f(x) = A \exp\left(\frac{(x -\mu)^2}{2\tilde\sigma^2}\right)
\end{equation}
for the sample size of N (red curve in Figure \ref{Fig:limb_stats}b). We then fitted a Gaussian distribution to the histogram of the observations (blue curve in Figure \ref{Fig:limb_stats}b) and find that the fitted parameters for the maximum $A$, mean $\mu$, and standard deviation $\tilde\sigma$ are almost identical to the theoretical values for this sample, see labels in Figure \ref{Fig:limb_stats}b. 

The obtained residual limb bin brightnesses are hence fully consistent with statistical random fluctuations. Although the mean of the fitted distribution is slightly positive ($\mu=0.08$), the only two outliers above 3 are negative and thus not indicative of emissions but would rather indicate absorption. 
No positive outliers beyond $I_{bin}/\sigma_{bin} = 3$ are observed. Statistically,  a value exceeding 3 would be expected for a sample with size N=414 with a probability of 60\%.

\subsection{Statistical tests with synthetic data}
\label{sec:synthetic}

We test the statistical robustness of our results by generating synthetic images of Europa’s emissions and applying the full analysis pipeline to them. To do this, we simulate statistical noise in each of the 23 model images by drawing random values from a Poisson distribution, using the model pixel value (in total counts) as the expected mean.

We then apply the same limb-bin analysis to the 23 synthetic images that we applied to the STIS images. The resulting distributions are again very similar to the expected Gaussian behavior. Figure \ref{Fig:SynthLyaImgs} shows one example set of 23 synthetic data images and can be compared to Figure \ref{Fig:limb_stats} (bottom). Furthermore, using 10,000 generated realizations of the full set of 23 images (one model per observation), we test the expected frequencies of obtaining outliers at various significance levels, which could otherwise be falsely interpreted as detections.

In the 10,000 samples of image sets, we find 6,790 3$\sigma$-outliers (i.e., an bin with a brightness that is $\geq$3$\sigma$ above the average limb brightness), corresponding to a probability of
\[
\frac{6{,}790}{10{,}000 \times 23 \times 18} = 1.6 \times 10^{-3},
\]
which is close to the theoretical one-sided $3\sigma$ detection probability of $1.4 \times 10^{-3}$. This includes several outliers in an image set (or even in one image). In 4,970 image sets, i.e. in roughly half of the total 10,000 image sets, there was no 3$\sigma$-outlier, as is the case for our STIS image set. In the other 5,030 image sets, at least one image contained at least one bin with a 3$\sigma$-outlier, but in many cases there were several outliers in one image set thus leading to the 6,790 total 3$\sigma$-outliers mentioned above.

A 4$\sigma$-outlier is found in only 178 image sets (as compared to the 5,030 image sets for 3$\sigma$), corresponding to a probability of $4.3\times10^{-5}$, which is again close to but slightly higher than the theoretical value of $3.2\times10^{-5}$. Thus, a 4$\sigma$ outlier would indeed be considered significant, but is only found in our STIS observation image in December 2012 if the disk position is assumed as in \cite{roth14-science}.

We also test the effect of positional offsets by shifting the synthetic noisy image along $x$ and/or $y$ and performing the analysis using the non-shifted model image as the reference. Such an offset introduces a systematic error and effectively broadens the distribution shown in Figure~\ref{Fig:limb_stats} (bottom), increasing the frequency of bin brightnesses at large $\sigma$ (as expected). For a shift by only one pixel in one direction, the frequency of a 4$\sigma$-outlier increases by a factor of $\sim$3. For a shift of one pixel in both $x$ and $y$, a 4$\sigma$-outlier occurs about 6 times more often. When shifting two pixels in $x$ and one pixel in $y$ -- the difference between this analysis and the \cite{roth14-science} analysis -- the factor is already $\sim$20, and there is a 35\% chance of obtaining one 4$\sigma$-outlier in one set of 23 images. This means that the 4.2$\sigma$ outlier derived when assuming the -- possibly false -- disk position used in \cite{roth14-science} is well consistent with statistical variations. In other words, the disk offset is a reasonable explanation for this outlier. 

We note that this test is not fully self-consistent, as the model image would slightly change under such a shift as well, because it is adjusted in brightness to best match the observation at the initially assumed position. However, it roughly quantifies how an incorrectly assumed position affects the probabilities of obtaining a false detection.

\subsection{Comparison to previous results for visit 3}
\label{sec:vis3}

While our limb-analysis results are very similar to those of \cite{roth14-science} and \cite{roth14-apocenter} for visits 1, 2, 4, and 5, we find no emission surplus in visit~3, in contrast to the previous study. The data processing and analysis presented here differ in several aspects from the approach used in \cite{roth14-science}. We use a slightly updated description of the surface reflection, accounting for the phase-angle dependence following \cite{oren94}, and we apply a reduced contrast to the inverted visible map. However, these differences are minor and have little impact on the off-disk signal in the limb-bin region.

More importantly, the earlier analysis did not account for emissions from the H exosphere, which was detected only later by \cite{roth17-europa}. The H exosphere remained undetected in the first five STIS datasets (visits 1-5), partly because the exospheric signal was comparatively low: Europa's radial velocity happened to be particularly small during these observations (see Table~\ref{tab:res} and Figure~\ref{Fig:H_exo}e). From visit 6 on, the exosphere signal was higher and clearly detectable. Nevertheless, even in visit~3, the peak brightness of the H exosphere at the limb was 145~R (value from Table~\ref{tab:res} multiplied by $\pi$), and the average brightness in the region between 1.0 and 1.25~R$_{\mathrm{Eu}}$ was 130~R. 

In the previous analysis, the only considered source of off-disk emissions above the constant background was the spread signal from on-disk solar reflection, which contributes on average up to only about 50~R in the limb region. When the H exosphere signal is taken into account, a higher intrinsic emission is expected in the limb region. This naturally reduces the residual brightness there. More importantly, it influences the determination of Europa’s disk position in the data, which turns out to be the crucial parameter, as it allows for homogeneously enhanced emissions above the limb.  

\begin{figure*}
   \centering
   \includegraphics[width=0.85\textwidth]{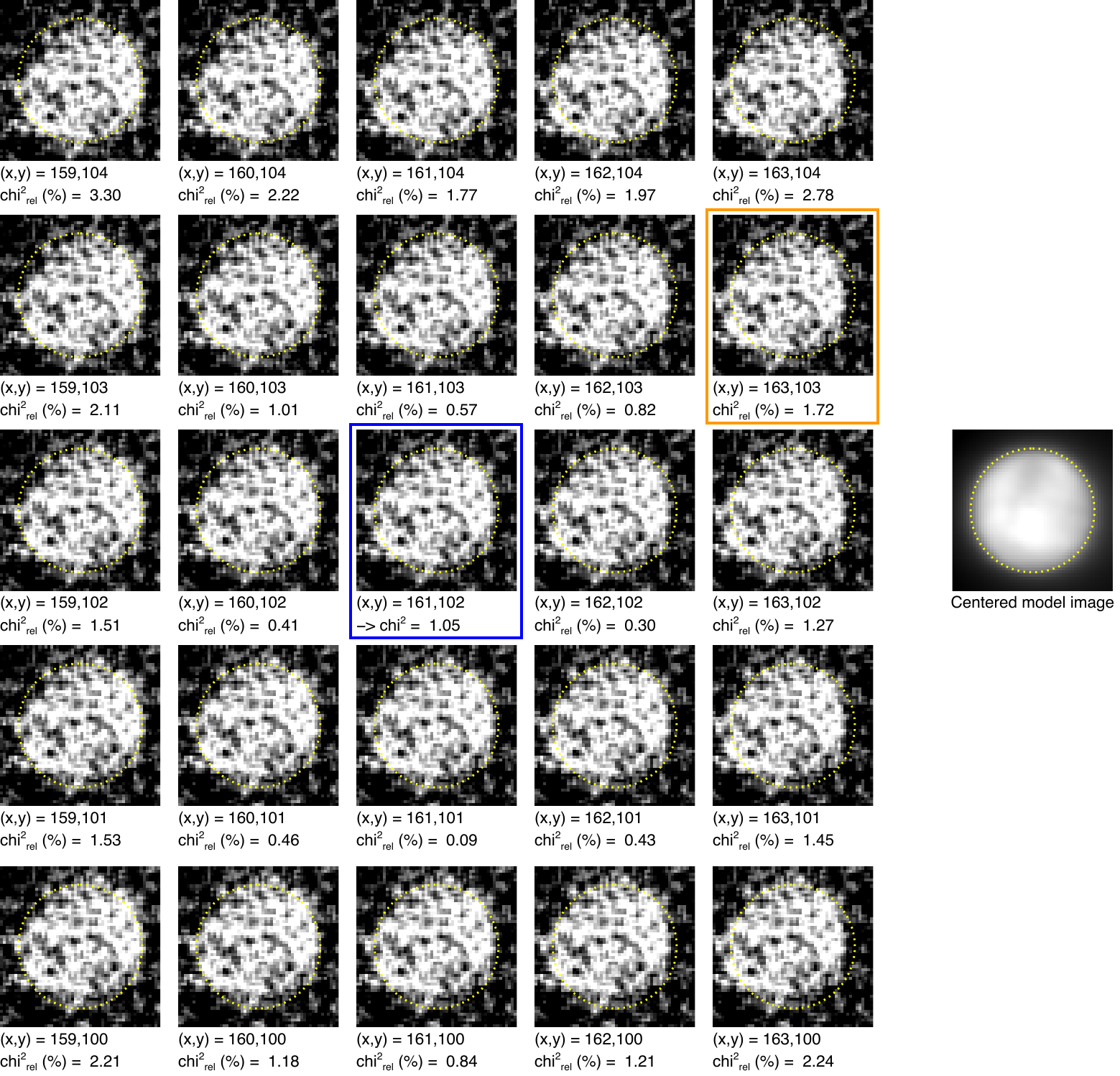}
   \caption{Overview of the disk positions around the best-fit position for visit 3 shown in the center with blue frame. The ($x,y$) position of the central pixel is given below in parenthesis and the corresponding disk limb is shown with the dotted yellow circle. Images are identical otherwise and all scaled and smoothed with 3x3 boxcar filter to enhance visibility. The disk position assumed in \cite{roth14-science} is indicted by the orange frame. The reduced chi-squared for all pixels is shown for the best-fit position. For all other positions the relative decrease of chi-squared value is shown in percentage of the best-fit value. The image from the forward model used to find the best-fit position is shown to the right. Note that the images are shown in detector frame and not rotated, unlike in Figure \ref{Fig:ResLyaImgs}.}
    \label{Fig:vis3_pos}%
    \end{figure*}

Figure~\ref{Fig:vis3_pos} presents different assumed positions of Europa's disk and the resulting chi-squared values for the comparison between the data and model images. The best-fit position (with a minimum chi-squared of $\chi^2 = 1.05$) is shown in the center with a blue frame. For the other positions, the relative increase in $\chi^2$ is shown as a percentage of the minimum value, since the absolute differences are small. The position assumed in \cite{roth14-science} is indicated by the orange frame. Using the algorithm described in Section~\ref{sec:mod}, the best-fit position differs from that used in \cite{roth14-science} by 2 pixels in the $x$-direction and 1 pixel in the $y$-direction. The crucial difference is the 2-pixel shift in $x$, as this moves the bright pixel patch in the lower-left quadrant either off or onto the disk (note that the images in Fig.~\ref{Fig:vis3_pos} are shown in the detector frame and are not rotated to North-up).

In \cite{roth14-science}, we assumed that the lower part of the disk (in the detector frame) was affected by the anomaly (i.e., the localized auroral emissions) and therefore used only the upper part of the disk to determine the central $x$ pixel. The details of the position-determination method are provided in the Supplementary Material of \cite{roth14-science}, in particular their Figure~S6. In the present analysis, we use the full image to determine the position using the comparison with a 2d model image, without assuming the presence of an anomaly \emph{a priori}. The contribution from the H exosphere considered here also influences the position search, as it accounts for relatively homogeneous signals above the limb more accurately. Despite providing a worse agreement with our model image, the disk position derived in \cite{roth14-science} still provides reasonable agreement with the data and cannot be excluded to be correct , given the noise level in the data.  

\begin{figure}
   \centering
   \includegraphics[width=0.49\textwidth]{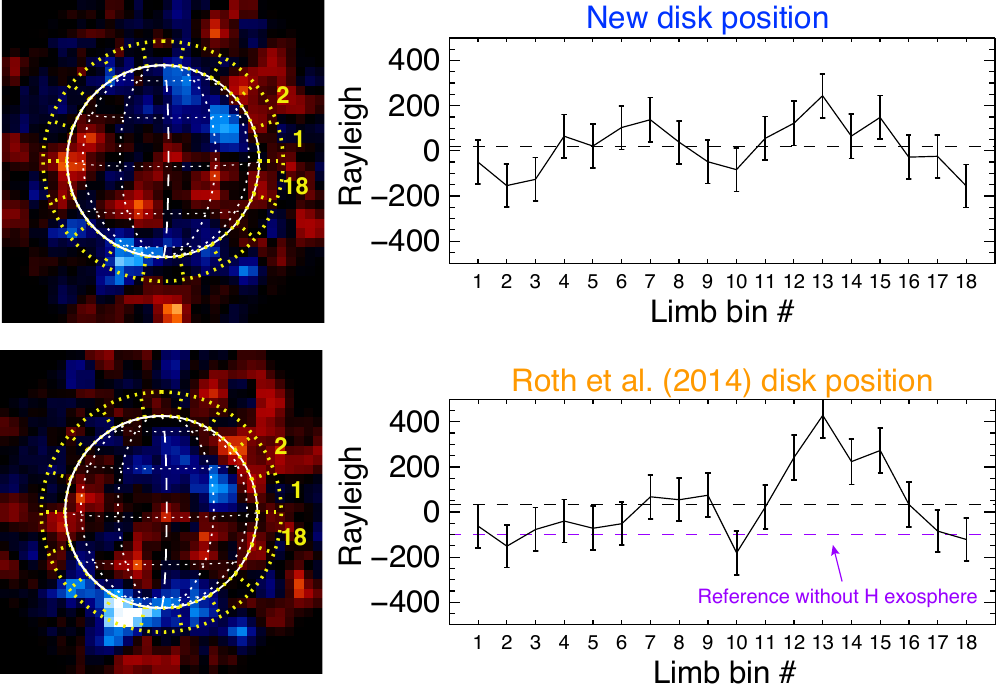}
   \caption{Comparison of limb bin analysis for visit 3 with the new disk position (top) and the disk position (bottom) used by \cite{roth14-science}. The left panels show the residual Ly$\alpha$ emission and the top image is identical to Figure \ref{Fig:ResLyaImgs}, panel V3. The 18 limb bin regions are shown by the yellow dotted lines. The right panels show the average residual brightness in each bin with error and the average brightness of all bins (black dashed). The purple dashed line shows a reference average brightness when an H-exosphere brightness of 130~R is subtracted.}
    \label{Fig:vis3_comp}%
    \end{figure}

Figure~\ref{Fig:vis3_comp} shows the limb-bin analysis for both positions. With the new position (top panel), the residual brightness in the bins is symmetrically distributed around the average limb brightness (black dashed line), ranging from approximately $-200$~R to $+200$~R. Bin~13 is the brightest, with $I_{\mathrm{bin}} = 244(\pm 98)$~R, corresponding to a brightness signal-to-noise ratio of 2.5 (see Figure~\ref{Fig:limb_stats}a for visit 3). This is the same bin for which the maximum brightness was inferred in \cite{roth14-science}, where a value of $I_{\mathrm{bin}} = 604(\pm 140)$~R was reported.

With the disk position from 2014, the bin brightnesses on one side of the disk (bins 7--15, except 10) are positive, while they are negative on the opposite side (bins 16--18 and 1--6). On the positive side, bin~13 is again the brightest, with $I_{\mathrm{bin}} = 430(\pm 102)$~R. The difference from the value reported for bin~13 in \cite{roth14-science}, $604(\pm 140)$~R, of 174~R can be partly explained by the subtraction of the H-exosphere brightness in the new analysis, which has an average value of 130~R in the limb region. When using a reference brightness that is lower by 130~R for the limb-bin measurements, together with the old disk position (dashed purple line in Figure~\ref{Fig:vis3_comp}, bottom), the resulting profile becomes very similar to the profile shown in Figure~3 of \cite{roth14-apocenter}.

Even without correcting the reference brightness, the brightness of bin~13, $I_{\mathrm{bin}} = 430(\pm 102)$~R, corresponds to $I_{\mathrm{bin}}/\sigma_{\mathrm{bin}} = 4.2$. This would imply a detection with a slightly higher significance than the value of 4.0 reported in \cite{roth14-science}. The absolute brightness of the maximum bin is lower (430~R vs. 604~R), but the brightness significance is higher because the earlier study used a more conservative error estimate, leading to a larger uncertainty of 140~R compared to 102~R here. In the previous error analysis, we included systematic uncertainties when subtracting the background and disk reflected contributions, while we here do not consider systematic uncertainties in the model fit and when subtracting the modeled contributions. The independent statistical test using the full limb-bin sample of all currently available images (Section~\ref{sec:limbbin} and Figure~\ref{Fig:limb_stats}b) supports the robustness of the uncertainties derived in the present analysis. Thus, when adopting the previous position of Europa's disk on the detector, an emission surplus is detected with a significance similar to (and slightly higher than) that found in the earlier study.

\section{Discussion}

\subsection{Updates in processing and analysis}

We have presented an analysis of the complete set of the Ly$\alpha$ signal in the HST/STIS FUV G140L 52"x2" observations of Europa in sunlight. For this analysis, we build a forward model for the contributions from surface reflection and H exosphere resonant scattering of the solar Ly$\alpha$ flux as well as the emissions from the Earth exosphere and interplanetary medium. The forward model with its parametrized individual contributions is then fitted simultaneously to the observations to obtain Ly$\alpha$ albedo, H exosphere brightness, and fore- and background levels. In previous studies of Europa's atmosphere and surface with the Ly$\alpha$ STIS data \citep{roth14-science,roth14-apocenter,becker18}, we had not considered H exosphere contributions and independently corrected for fore- and background emissions. In addition, here we considered the surface illumination angle using Oren-Nayar reflectance model while previously we assumed a uniform disk. We also used an adjusted contrast map for the surface FUV reflection (adjusted inverted visible maps). Finally, possible inhomogeneities near the limb from potential plumes are constrained by characterizing differences between data and forward model images. 

\subsection{Europa's H exosphere: Density, temperature, and production and loss rates}

The derived vertical H column densities in Europa’s exosphere of \(1.4\times10^{12}\,\mathrm{cm^{-2}}\) (2012--2015) and \(1.1\times10^{12}\,\mathrm{cm^{-2}}\) (2018--2020) are \(4\text{--}5\) times higher than the densities reported by \citet{roth17-europa} based on the attenuation of Jupiter's daylow in the Europa exosphere in transit. However, the \citet{roth17-europa} densities need to be corrected and after the correction our results are actually in agreement: \citet{roth17-europa} underestimated the H density due to an overestimation of the effective cross section for the attenuation of the Jovian background dayglow, as pointed out in \citet{Roth2023}. When correcting the cross section to an about 5 times lower value \citep{Roth2023}, the inferred H density from the data of \citet{roth17-europa} is higher by a factor of about five. Therefore, the values reported here are consistent with the H exosphere detection in transit \citep{roth17-europa} when the appropriate attenuation cross section is used.
 
We have separately fitted the two periods (2012-2015 vs 2018-2020) as the resulting densities appeared to converge to systematically different values. The resulting fitted densities and their relatively small uncertainties suggest that Europa's H exosphere has indeed a different density in the two periods (Table \ref{tab:extinct_fit}), with a $\sim$25\% decrease in density from the first period to the second. This is a result independent of the attenuation of Earth and derived primarily from the observations taken at large Doppler shifts, which almost all happened either in late 2014/early 2015, or in 2018. The measurements taken at radial velocities lower than 10 km/s (from 1999, early 2014, 2012, 2019 and 2020, see Figure \ref{Fig:H_exo}, right column) are strongly affected by the attenuation at Earth and possible differences in Europa H density might be masked by this. The H exosphere signal is not detected during visit 24 likely because the expected weak exosphere signal (green curve in Figure \ref{Fig:H_exo}f) could not be differentiated from the background trend in our fitting algorithm.  

The best-fit temperatures for Europa's H exosphere were 1000~K and 2100~K for the two periods, respectively. The temperature is constrained primarily by observations taken at Doppler shifts between 3~km\,s\(^{-1}\) and 10~km\,s\(^{-1}\), where the transition from strong attenuation in Earth's exosphere to full transmission occurs. The temperature derived for the first period is more reliable because several observations were obtained within this radial-velocity transition range only a few months apart, making the simultaneous fit a more reliable measure. In the second period, only the visit~28 observation had an intermediate Doppler shift, which in addition carried a large uncertainty and was taken a year after the high--Doppler-shift observations so that other factors might have changed.

Using the density and temperature, we estimate the hydrogen source rate, $Q_H$, again assuming a cometary-like escaping exosphere. In this framework, $Q_H$ is given by
\begin{equation}
    Q_H = 4\pi \, n_{\mathrm{H},0} R_E^2\,v = 4\pi \, N_{\mathrm H} R_E\,v \quad ,
    \label{eq:Q_H}
\end{equation}
where $v$ is the outward velocity of H. We adopt the most probable thermal velocity for an ideal gas,
\begin{equation}
    v = \sqrt{\frac{2k_B T_H}{m_H}} \quad .
\end{equation}
For a temperature of $T_H = 1000$~K, derived from the 2012--2015 datasets, this corresponds to $v = 4.1$~km~s$^{-1}$. Using a column density of $1.4 \times 10^{12}$~cm$^{-2}$, we obtain a nominal production rate of $Q_H = 1.1 \times 10^{27}$~H~s$^{-1}$.

Assuming a Maxwellian velocity distribution for the upward velocity component, with a characteristic speed of 4.1~km~s$^{-1}$, approximately 93\% of the H atoms exceed the Hill escape velocity ($v_{\mathrm{esc,Hill}} = 1.96$~km~s$^{-1}$) and are therefore able to escape. This implies an H escape rate from Europa of $\sim 1.0 \times 10^{27}$~H~s$^{-1}$. The remaining, non-escaping H atoms re-impact the surface and undergo chemical reactions.

Ionization can be neglected for this estimate. It takes roughly 5000~s for an H atom to travel from the surface to Europa's Hill sphere (a distance of $\sim 12{,}000$~km, or $\sim 8$ Europa radii) at a velocity of 4.1~km/s. Electron-impact ionization rates are well below $10^{-5}\ \mathrm{s}^{-1}$, so the H atoms escape before being ionized. Rates for charge exchange with ambient ions are again an order of magnitude lower than electron-impact rates, and thus charge exchange is even more negligible \citep{smith19}.

With the lower density derived from the 2018 observations ($N_H = 1.1\times10^{12}$~cm$^{-2}$), the source rate is also lower by 25\%. A higher H temperature and thus $v$ on the other hand increases the source rate but even more the loss rate. With $T_H = 2100$~K, $v$ is 5.7~km/s and the escaping fraction of H is 96\%. At $T_H = 5000$~K, 97\% of the H escapes directly.

It is important to emphasize that the assumption of a single temperature describing the effective Ly$\alpha$ line width and thus attenuation might not capture the actual velocity distribution of the H atoms. Line-resolved measurements of atomic sodium (Na) emissions at Europa showed variation in line width and suggested an increase in Na temperature with distance from Europa \citep{Lovett2025}. The strongest signal in our data is from within 1~$R_E$ altitude, and our data would not be sensitive to, for example, an increase in temperature at higher altitudes. For example, the model of \cite{smyth06} yields an H temperature of 1000~K near the surface, which increases to about 3000~K at 1000~km altitude. The authors explain the temperature profile with a mixture of colder surface-sourced H atoms and hotter H atoms produced by molecular dissociation as well as collisions with the thermal O$_2$-atmosphere. Our results for the H temperature are consistent with this profile from \cite{smyth06}. We discuss the expected densities and temperatures from different sources for H in the following subsection. However, we note already that significantly higher H temperatures are expected from dissociation of atmospheric molecules to produce atomic H due to excess energies, which would suggest even higher escape rates than estimated with the derived H temperatures.

The obtained column densities are very similar to the H column densities measured at Ganymede \citep[(1--2)\(\times10^{12}\,\mathrm{cm^{-2}}\),][]{barth97,alday17,Roth2023} and Callisto \citep[\(\sim1\times10^{12}\,\mathrm{cm^{-2}}\),][]{barth1997b,roth17-callisto}. Given these comparable column densities and the fact that the radii of Ganymede and Callisto are 69\% and 55\% larger than Europa's radius, the H number density at the surface, $n_{\mathrm{H},0}$, is higher at Europa than at the other two moons assuming the same $1/r^2$ density profile in all cases. The larger radii of Ganymede and Callisto, however, lead to larger total content of the global H exosphere and therefore larger source rates $Q_H$ (Equation \ref{eq:Q_H}). On the other hand, Europa has a lower escape velocity than Ganymede and Callisto and therefore the escaping fraction of the H exosphere will be higher at Europa (in other words, higher gravity prevents escape of H more at Ganymede and Callisto). Since Europa orbits closer to Jupiter than Ganymede and Callisto, the same amount of escaping H will lead to higher densities along Europa's smaller orbit. 

\subsection{Europa's H exosphere: Source mechanisms and implications}

We now examine whether the derived constraints on H exosphere density (considering also the difference between the two periods), temperature, and production and loss rates are consistent with potential source mechanisms for the H atoms at Europa. Possible sources for H are:
\begin{itemize}
    \item Dissociation of H-bearing molecules in the atmosphere, primarily H$_2$ or H$_2$O, driven by either UV photons or electron impact. 
    \item Dissociation of the surface ice followed by direct release of H atoms. The can be induced by charged particle sputtering \citep[e.g.,][]{BarNun1985} or photolysis \citep{Johnson1997}.
\end{itemize}
We first discuss the dissociation of atmospheric molecules, before we briefly investigate a direct surface source.

Photodissociation of H$_2$ or H$_2$O is unlikely to be the dominant source of atomic hydrogen. First, electron-impact dissociation has been shown to be significantly more efficient than photodissociation at Europa \citep{shematovich05,smyth06,szalay2024}. In addition, H atoms produced by photodissociation of H$_2$ or H$_2$O acquire excess energies of $\sim$1~eV or more \citep[e.g., Table~2 in][]{marconi07}, implying temperatures $T \gtrsim 10^4$~K, well above the upper limit derived here. Furthermore, although the solar UV flux decreased from 2015 to 2018 -- potentially leading to a reduced photodissociation rate -- the magnitude of this change was smaller \citep{woods05} than the observed $\sim$25\% difference in H density. Moreover, the present short-term solar UV variability ($\sim$20\%) is not reflected in the inferred H densities at Europa.

Electron-impact dissociation of H$_2$ or H$_2$O is expected to be more efficient at Europa than photodissociation. The energies of the H atoms produced by electron-impact dissociation are not well constrained, and the inferred H temperature therefore cannot be used to validate or invalidate this process. Nevertheless, electron impact is likely still insufficient to account for the required H production rate.

Because the abundance of H$_2$O is likely significantly lower than that of H$_2$, we focus on H$_2$ for the following estimates. To sustain an H production rate of $Q_H = 1.1 \times 10^{27}$ H s$^{-1}$ via dissociation of H$_2$, an H$_2$ production rate (e.g., from radiolysis) of $\tfrac{1}{2}Q_H = 5.5 \times 10^{26}$ H$_2$ s$^{-1}$ would be required to compensate for the loss to H. In addition, an independent H$_2$ source rate of $4.5(\pm2.4) \times 10^{26}$ H$_2$ s$^{-1}$ is needed to explain the observed H$_2^+$ pickup ions, assuming escape of neutral H$_2$ followed by ionization in a torus \citep{szalay2024}.

Because dissociative recombination of H$_2^+$ is negligible \citep{szalay2024,vanderZande1996}, dissociation into H versus escape and ionization can be considered two independent pathways for H$_2$. The total required H$_2$ source rate to sustain the H exosphere and explain magnetospheric H$_2^+$ pick-up ion densities would then be
\begin{equation*}
    5.5 \times 10^{26} \;\mathrm{H}_2\;\mathrm{s}^{-1} + 4.5 \times 10^{26} \;\mathrm{H}_2 \;\mathrm{s}^{-1} = 1.0 \times 10^{27} \;\mathrm{H}_2 \;\mathrm{s}^{-1} \quad.
\end{equation*} Under this scenario, 55\% of the H$_2$ would need to be dissociated to form H, while only 45\% would be ionized to form H$_2^+$. This partitioning is unlikely, because H$_2$ escape occurs on a timescale of $\sim10^4$~s, whereas the dissociation time scale (inverse rate) is on the order of $10^{6}$~s. Consequently, only of order 1\% (not 55\%) of the H$_2$ is expected to dissociate before escaping.

In addition, the electron conditions depend on the local plasma environment, which is modulated primarily on the synodic rotation period of Jupiter of 11.2~h \citep{roth16-eur}. The visits were intentionally scheduled at different plasma environments (see System-III longitude in Table \ref{tab:obs}) and there is no systematic pattern that could explain the difference in H density between the two periods that we inferred here. Europa is deep inside the magnetosphere and seasonal changes or changes in the solar wind do not affect the local plasma environment.

Instead, the difference in density between the two observing periods is more likely explained by changes in the source rates of the parent molecules. The radiolysis yield of surface ice induced by charged-particle impacts has been shown to depend strongly on surface temperature, increasing approximately exponentially with temperature \citep[e.g.,][]{shi95,johnson09,raut2013}. Using the temperature dependence reported by \citet{shi95} for sputtering by O$^+$ ions (see their Fig.~2), we find that an increase in surface temperature from 100~K to 103~K --- a 3\% change near Europa’s average surface temperature \citep{spencer99} --- would result in a $\sim$28\% increase in radiolysis yield, comparable to the observed difference in exospheric density. Such a temperature-driven change would directly affect the production rate of H$_2$ as parent species supplying the H exosphere.

However, it is unclear whether the temperature variations of the surface (and the corresponding changes in radiolytic production yield) are indeed sufficiently large to explain the observed density difference. Seasonal surface temperature variations are poorly constrained. During the earlier period, Europa's distance from the Sun was $\sim$5.3~AU, compared to $\sim$5.4~AU in the later period; this change corresponds to only a 3\% difference in solar irradiance at Europa. Variations in solar irradiance due to the solar cycle are negligible in this context, because the power in the visible and infrared portions of the solar spectrum remains nearly constant \citep{Woods2022}. Modeling of Europa's surface temperature \cite{ashkenazy2019} suggests maximum temperature changes of only $\sim$5~K over a full Jovian year, despite $\sim$20\% changes in solar flux at Europa between aphelion and perihelion. If radiolytic production yields indeed change significantly due to surface temperature variations, all radiolytically produced atmospheric species -- including O$_2$ -- would exhibit strong seasonal variability.

A hint that seasonal variation might indeed happen is provided by the 1999 HST observation. In 1999, Europa was closest to the Sun at 4.96~AU. The H density derived here, although affected by the Earth exosphere, is clearly higher than the best-fit profiles suggest (gray data points vs solid green lines in Figure \ref{Fig:H_exo}f,g). Similarly, the oxygen aurora measured in the same 1999 observation and produced primarily by dissociative excitation of O$_2$ was by far the brightest \citep[][their figure 6]{roth16-eur}. 

A similarly strong temperature dependence is given for ice sublimation as a source for atmospheric H$_2$O \citep[e.g.,][]{leblanc17}. However, production of H from purely sublimated H$_2$O is insufficient due to the low expected H$_2$O abundance, and this would not produce a global, uniform H abundance. 

Finally, direct production of H from erosion of the icy surface might be an alternative explanation. The low inferred H temperatures constitute a strong argument for this source, because production via dissociation of atmospheric molecules likely lead to temperatures at least an order of magnitude higher (for both photodissociation and electron impact dissociation). It is, however, difficult to assess the efficiency of surface H sources. 

There is only one published study on direct sputtering of atomic H off the surface from charged particle irradiation by \cite{BarNun1985}. Whether the yield would be sufficient to explain the H source rate obtained, and what H temperature would be expected is unclear. \cite{BarNun1985} mention that the yield they find is independent of the ice temperature, which means the density difference between the two periods could hardly be explained by the direct sputtering source (unless the sputtering flux changes). Incident UV photons can also cause direct dissociation of the water ice in the surface via photolysis, leading to production of atomic H \citep[e.g.,][]{Johnson1997}. The efficiency and characteristics of such photolysis-induced desorption of H is yet also not known. Similar to photodissociation of atmospheric molecules, changes in solar UV flux between the two periods might partially explain the found H exosphere variability (see discussion above). However, whether it can account for the quantitative change of 25\% can not be estimated due to lack of knowledge about the process.     

Regardless of whether H is produced from the surface directly or from atmospheric hydrogen-bearing molecules, the total hydrogen source (H plus H$_2$) must be roughly a factor of $\sim 2$ higher (in mass or atoms) than recent estimates by \citet{szalay2022,szalay2024} based on H$_2$ only. Assuming that O$_2$ is produced at the stoichiometric ratio, this also implies an O$_2$ production rate about twice as high as in \citet{szalay2024}, bringing it closer to the source rates commonly assumed in atmospheric simulation studies \citep[e.g.,][]{shematovich05,smyth06,plainaki13}.

The inferred H density and escape rate are about an order of magnitude higher than assumed in modeling studies of the atmosphere and escape to a neutral torus \citep{smyth06,smith19}. With higher values for density and escape, atomic H may possibly become the most abundant species in Europa's neutral torus and thereby relevant for the interaction with the charged particles \citep{mauk03,lagg03}. 

\subsection{Earth exosphere}

We also included the density of the Earth’s H exosphere as a fit parameter and obtained column densities (Table \ref{tab:extinct_fit}) within the range expected from exosphere models, in particular from the NRLMSIS 2.0 atmosphere model \citep{Emmert2020}. For our best-fit results, we find Earth-exosphere H column densities of $9.9 \times 10^{12}$ cm$^{-2}$ and $3.9 \times 10^{13}$ cm$^{-2}$ for the two periods 2012–2015 and 2018–2020, respectively. The difference is likely connected to the variation of Earth's exosphere with the solar cycle. 

The approximately four-times higher column density near solar minimum can be explained by the enhanced H abundance near the exobase -- in the lowest part of the exosphere -- reported in previous studies \citep{Nossal2012, Waldrop2013}. This elevated H abundance may be due to dynamical production and upward transport at low solar activity \citep{Qian2018}. 

During solar maximum, the temperature in the upper atmosphere increases, pushing more H to higher altitudes and into the far-extended exosphere. There, the correlation between H abundance and solar activity appears to be inverted compared to the region near the exobase, with more H observed during solar maximum \citep{Zoennchen2024}. However, the attenuation of the source signal occurs in the optically thick H exosphere above the exobase, i.e., within the first few hundred kilometers above Hubble ($\sim$540 km). There, the H density is higher during solar minimum, consistent with our results.

\subsection{Local inhomogeneities}

We did not identify any outliers in any of the 23 residual emission images. The distribution of the residual brightness in the limb bins is fully consistent with purely statistical variations.

Notably, we also do not find an emission surplus in the image from visit 3 taken in December 2012, in contrast to our earlier analysis and interpretation \citep{roth14-science}. The primary reason for the different results is a difference in the assumed position of Europa on the detector. The importance of this a priori unknown position is crucial in the analysis of HST observations, as previously demonstrated by \cite{giono20} for STIS transit FUV filter observations of Europa.

The other main difference is the inclusion of an H exosphere signal from Europa when correcting for other sources. In contrast to our earlier analysis \citep{roth14-science}, in which we assumed an anomaly in the data and used only a part of Europa's disk signal to determine its position, in this study we applied the same approach and algorithm to visit 3 as to all other visits. Therefore, the analysis presented here is preferred and considered more impartial. This means that the HST/STIS Ly$\alpha$ auroral observations do not provide evidence for a localized abundance of H$_2$O from outgassing. The conclusions on a stable H$_2$O atmosphere above the sunlit trailing hemisphere from \cite{Roth2021-Eur} are not affected by our results here as they were derived only from OI1304~Å and OI1356~Å images (and not Ly$\alpha$).   

\cite{roth14-science} discussed several lines of supporting evidence for the detection of localized H$_2$O aurora. In particular, an excess of OI1304~Å emission was reported at the same location as the Lyman-$\alpha$ excess. In addition, the persistence of the above-limb Lyman-$\alpha$ signal as well as an apparent correlation between brightness variations and the plasma environment were emphasized. Although these findings were never considered to constitute evidence on their own, we briefly discuss them in light of the new results for completeness.

The OI1304~Å excess was detected at only a 2.4$\sigma$ significance level in \cite{roth14-science} and was hence not a statistically significant outlier itself. With the revised disk location (Figure \ref{Fig:Vis3oxygen}), there is still a some increased OI1304~Å brightness left of the south pole, but the corresponding bin \#13 is well within the standard variation of brightness around the limb and other limb bins have higher brightness.   

The potential correlation to the plasma environment is still seen but the persistency of an emission surplus near the south pole throughout the visit disappeared: A Ly$\alpha$ excess in the south polar bin is observed during orbits 3 and 4 of the five HST observing orbits, while no excess is detected in orbits 1, 2, and 5 (for comparison see Figure 2, panels F–J, in \cite{roth14-science}). Thus, although the excess in the two orbits still coincides with plasma sheet crossings -- and therefore with the highest ambient plasma densities, where enhanced H$_2$O auroral excitation would be expected -- the lack of a statistically significant detection means that this does not constitute evidence for an outlier or local H$_2$O aurora.

Studies of potential plumes at Europa following the initial detection by \cite{roth14-science} have built on those results to varying degrees. The potential plume signals found in STIS FUV filter images \citep{sparks16} was shown to be possibly explained by statistical fluctuations and a misalignment of the assumed disk position (similar to the results here) by \cite{giono20}. \cite{jia18} presented plasma fluid simulations that reproduce observed plasma wave and magnetic field signatures measured by the Galileo spacecraft during flyby E12, assuming the presence of a localized, jet-like plume of neutral gas. Recently, \cite{Paterson2026} showed that data from the Galileo Plasma Instrumentation does not confirm the density enhancement derived from the Galileo wave data by \cite{jia18}, discounting a plume encounter during flyby E12 according to the authors.

\cite{paganini2020} reported a single detection of H$_2$O vapor emission above the leading hemisphere during an infrared observing campaign with the Keck telescope spanning multiple dates; the study interpreted this result as indicative of intermittent activity. Subsequent non-detections in infrared observations by the James Webb Space Telescope \citep{villanueva2023} are consistent with the inferred infrequent nature of such events. At present, direct measurements of water-molecule emissions -- particularly via infrared observations and ideally with spatial resolution of the source -- likely constitute the most sensitive remote sensing observations for placing constraints on water plume activity at Europa. After arrival in 2030, NASA's Europa Clipper will systematically and globally constrain plume activity at Europa with its diverse instrument suite \citep{Pappalardo2024,roth2025}. 
 
\section{Summary}

We carried out a comprehensive analysis of Ly$\alpha$ observations of Europa obtained in 1999 and between 2012 and 2020. Our analysis includes forward modeling of all known contributions to the Ly$\alpha$ signal. 
%In this model, the surface-reflection component constrains the Ly$\alpha$ albedo, which peaks near $200^\circ$~W longitude.
The study focuses on deriving constraints on Europa’s hydrogen exosphere and on performing a systematic search for localized emission outliers. The key findings regarding the H exosphere can be summarized as follows:
\begin{itemize}[label=$\bullet$]
\item H exosphere emissions are detected in all but one observation, with varying brightness.
\item Ly$\alpha$ illumination from interplanetary H can enhance the flux scatterable by Europa's H exosphere up to 20\%.
\item A primary factor controlling the observed exosphere brightness is attenuation of the Europa signal in Earth’s atmosphere, which is significant when the line-of-sight velocity of Europa relative to Earth is less than 10~km s$^{-1}$.
\item Europa exospheric H column densities of $1.4\times10^{12}\ \mathrm{cm^{-2}}$ and $1.1\times10^{12}\ \mathrm{cm^{-2}}$ were derived for the 2014/2015 and 2018 observations, respectively.
\item An effective temperature of approximately 1000~K was inferred from the 2014/2015 observations, with an upper limit of 5100~K.
\item The inferred H source rate of $1.1 \times 10^{27}$~s$^{-1}$ is more than twice the H$_2$ source rate recently derived from H$_2^+$ pickup ions. This suggests that a significant fraction of the hydrogen produced by erosion of surface H$_2$O ice ultimately becomes atomic hydrogen.
\item The high H density and source rate, combined with the relatively low H temperature, are difficult to reconcile with the current understanding of possible source processes. The low temperatures seem to preclude a production from dissociation of molecular species in the atmosphere.
\end{itemize}

In addition to the findings on Europa’s exosphere, our analysis provides evidence that the H column density in Earth’s exosphere above HST’s altitude (550~km) is higher during solar minimum. This result can be tested by the recently launched NASA Carruthers Geocorona Observatory, which will investigate the H exosphere in detail \citep{waldrop2024nasa,JoshiP2024}.\\

The results of the search for localized emissions are summarized as follows:
\begin{itemize}[label=$\bullet$]
\item No statistically significant emission surplus is found in the 23 HST images of Europa obtained in sunlight.
\item The derived brightnesses significance in the limb bins are in good agreement with statistical fluctuations.
\item The difference from the previous results published in \cite{roth14-science} arises from differences in the assumed position of Europa on the detector and from the inclusion of the H exosphere in the analysis.
\item When adopting the same assumed position for Europa as in \cite{roth14-science} and neglecting the contribution from the H exosphere, an outlier would again be identified in the December 2012 observations. Using an updated error analysis, the derived significance of this outlier would be 4.2$\sigma$, slightly higher than the significance of the outlier reported in the previous study (4.0$\sigma$).
\end{itemize}

\begin{acknowledgements}
This research is based on observations made with the NASA/ESA Hubble Space Telescope obtained from the Space Telescope Science Institute, which is operated by the Association of Universities for Research in Astronomy, Inc., under NASA contract NAS 5–26555. These observations are associated with program(s) HST-GO-8224, HST-GO-13040, HST-GO/DD-13619, and HST-GO-13679.
LR acknowledges support by the Swedish National Space Agency through grants 2021-00153 and 2024-00112. SRCM is supported by the Swedish Research Council through Starting Grant 2025-04693. SJ is supported by Swedish National Space Agency grant 2020-00187. 

\end{acknowledgements}

\bibliographystyle{aa} % style aa.bst
\bibliography{bib2025} % your references Yourfile.bib

% \begin{thebibliography}{}

%   \bibitem[Baker(1966)]{baker} Baker, N. 1966,
%       in Stellar Evolution,
%       ed.\ R. F. Stein,\& A. G. W. Cameron
%       (Plenum, New York) 333

%\end{thebibliography}

\begin{appendix}%First appendix

\section{IPH Ly$\alpha$ brightness and map}
\label{sec:A_IPMmap} 

\begin{figure*}[hb]
\sidecaption
   \includegraphics[width=12cm]{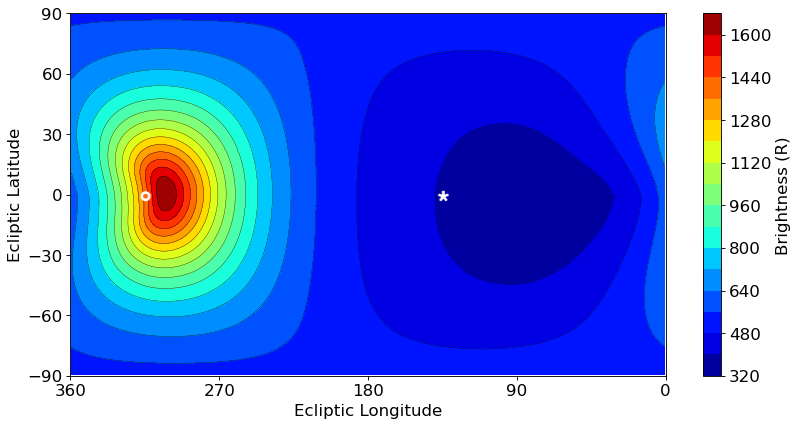}
   \caption{Sky map of the Ly$\alpha$ brightness seen from Europa for visit 12 (Jan 26, 2015) from interplanetary medium H scattering model \citep{Pryor2024}. The direction to the Sun is shown by the small white circle, the anti-Sun direction is shown by the white asterisk.}
\label{Fig:IPMmap}%
\end{figure*}

Figure \ref{Fig:IPMmap} shows an example map of the Ly$\alpha$ sky brightness produced by sunlight scattered by the IPH. The region of maximum emission is located near the Sun on the upwind side of the IPH flow. Very close to the Sun, the density of the IPH is depleted by charge-exchange processes; this depletion forms a cavity that extends downstream with the IPH flow, reducing the brightness on the downstream side.

The brightness from these modeled maps is used to estimate the scattering of IPH Ly$\alpha$ radiation by Europa’s H exosphere, in addition to the scattering of direct solar illumination, as described in Section \ref{sec:Hexo_col}. The $g$ factor is determined using a sky map of the relative radial velocity of the IPH and by calculating the spectral flux in all directions while accounting for Doppler shifts (see Figure \ref{Fig:lya_prof}a). 

For reference, the average brightness on the dayside and nightside hemispheres for this day were 790~R and 450~R, respectively. Integrating over each hemisphere, one yields total fluxes of $4.1\times10^8$~photons/cm$^2$/s for the dayside  and $2.3\times10^8$~photons/cm$^2$/s for the nightside. 

\section{Ly$\alpha$ albedo variation}
\label{sec:A_albedo} 

\begin{figure}[hb]
   \centering
   \includegraphics[width=0.49\textwidth]{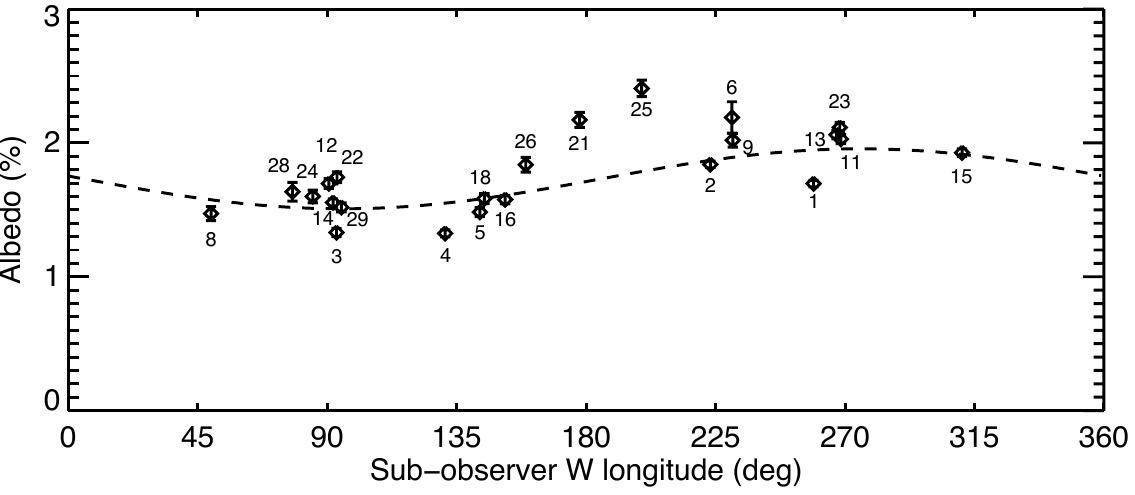}
   \caption{Ly$\alpha$ albedo values derived from the 23 visits and a fitted sinusoidal periodic variation.}
\label{Fig:albedo}%
\end{figure}

The derived Ly$\alpha$ albedo values (Table~\ref{tab:res}) confirm the previous results of \cite{becker18}. Our values are overall slightly higher because we use the Oren--Nayar reflectance model, which assumes reduced reflectance near the limb compared to a uniformly bright disk. The new 2018--2020 data fill a gap around 160--200$^\circ$ sub-observer west longitude (see Fig.~4A of \cite{becker18}), yielding the highest overall albedo at 200$^\circ$ (visit~25), see Figure \ref{Fig:albedo}. Fitting a sinusoidal variation to all albedo values, the fitted peak remains near 270$^\circ$, as in the previous study, but there also appears to be an additional increase towards the anti-Jovian hemisphere (180--200$^\circ$\,W), similar to what \cite{becker18} found for the albedo at 1335~\AA. 

\section{Example set of synthetic images }
\label{sec:A_SynthData}

\begin{figure*}[hb]
   \centering
   \includegraphics[width=0.9\textwidth]{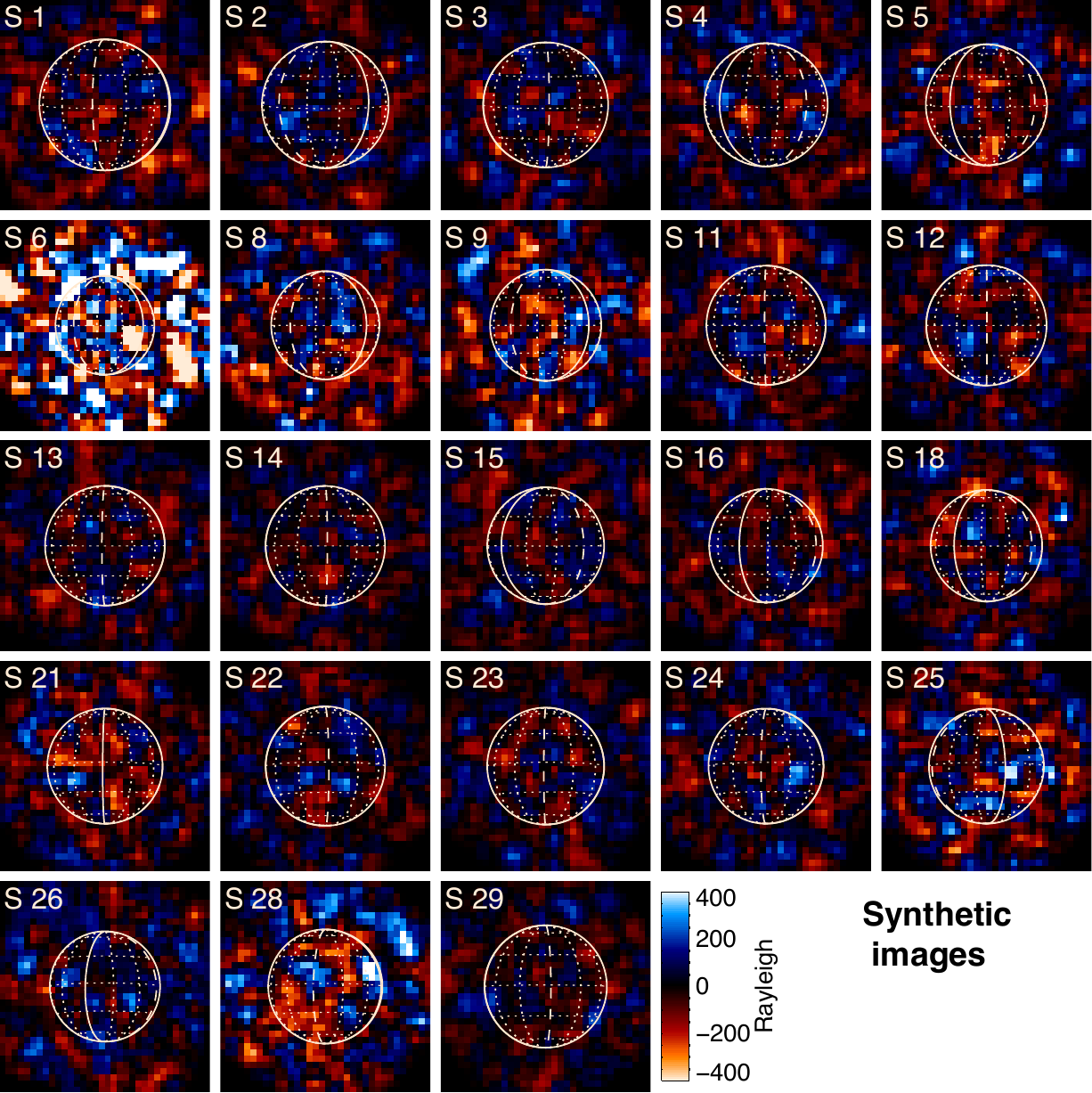}
   \caption{Synthetic Ly$\alpha$ images generated by adding Poisson noise to the forward model image for each of the 23 visits. The synthetic images are then processed and displayed in the same way as the STIS observation images shown in Figure \ref{Fig:ResLyaImgs}, including binning (2x2 detector pixels) and boxcar smoothing (3x3 binned pixels) for display. The noise level relates to the total counts in each image reflecting the noise in the observations.}
\label{Fig:SynthLyaImgs}%
\end{figure*}

Figure \ref{Fig:SynthLyaImgs} shows an example set of images based on synthetic noise data. 

\section{Oxygen images from visit 3}
\label{sec:A_OxygenVis3}

\begin{figure}[hb]
   \centering
   \includegraphics[width=0.5\textwidth]{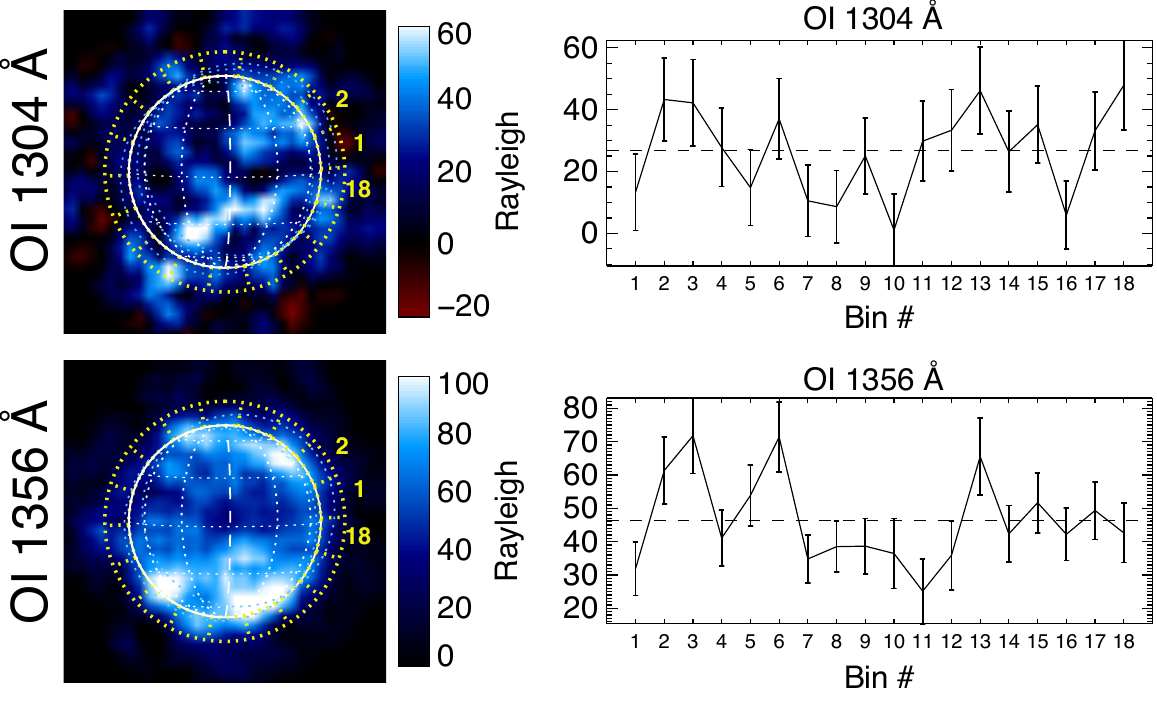}
   \caption{The oxygen aurora images centered at 1304~Å and 1356~Å from visit 3 using the new disk position and brightnesses in the 18 limb bins. For comparison see Figure 1 (panels L and O) and Figure 3 (C and E) in \cite{roth14-science}.}
\label{Fig:Vis3oxygen}%
\end{figure}

Figure \ref{Fig:Vis3oxygen} shows the oxygen aurora images with the new position of Europa and the limb bnin analysis. 

\end{appendix}

\label{LastPage}
\end{document}